\documentclass[preprint,12pt]{elsarticle}
\usepackage{amssymb}
\usepackage{amsthm}
\usepackage{pifont}
\usepackage{subcaption}
\usepackage{algorithm}
\usepackage{algpseudocode}
\usepackage{hyperref} 
\usepackage[shortcuts,acronym,nonumberlist,nogroupskip]{glossaries}
\usepackage{tabularray}
\usepackage{adjustbox}
\usepackage{nomencl}
\usepackage[table,xcdraw]{xcolor} 
\newcommand{\cmark}{\ding{51}}%
\makeglossaries 
\newacronym{am}{AM}{active material loss}
\newacronym{bm}{BM}{bucket model}
\newacronym{bms}{BMS}{battery management system}
\newacronym{bess}{BESS}{battery energy storage system}
\newacronym{da}{DA}{day-ahead}
\newacronym{der}{DER}{distributed energy resources}
\newacronym{dla}{DLA}{Direct Lookahead}
\newacronym{dso}{DSO}{distribution system operator}
\newacronym{ct}{CT}{continuous-time intra-day}
\newacronym{ems}{EMS}{energy management systems}
\newacronym{ema}{EMA}{energy management algorithm}
\newacronym{empc}{eMPC}{economic MPC}
\newacronym{ess}{ESS}{energy storage systems}
\newacronym{ecm}{ECM}{equivalent circuit model}
\newacronym{ev}{EV}{electric vehicle}
\newacronym{fom}{FOM}{full order model}
\newacronym{hess}{HESS}{hybrid energy storage system}
\newacronym{hp}{HP}{heat pump}
\newacronym{hv}{HV}{high voltage}
\newacronym{hvac}{HVAC}{heat, ventilation and air conditioning}
\newacronym{iso}{ISO}{independent system operator}
\newacronym{ida}{IDA}{intra-day auctions}
\newacronym{nmc}{NMC}{nickel manganese cobalt oxides}
\newacronym{mpc}{MPC}{Model Predictive Control}
\newacronym{mces}{MCES}{multicarrier energy systems}
\newacronym{ocv}{OCV}{open-circuit voltage} 
\newacronym{ocp}{OCP}{optimal control problem} 
\newacronym{pei}{PEI}{power electronic interface}
\newacronym{pbrom}{PBROM}{physics-based reduced order model}
\newacronym{rl}{RL}{reinforcement learning}
\newacronym{scm}{SCM}{series connected module}
\newacronym{sei}{SEI}{solid electrolyte interface}
\newacronym{sdp}{SPD}{sequential decision problem}
\newacronym{soc}{SOC}{State of Charge}
\newacronym{soh}{SOH}{State of Health}
\newacronym{spv}{SPV}{solar photovoltaics}
\newacronym{st}{ST}{solar thermal}
\newacronym{tess}{TESS}{thermal energy storage system}
\newacronym{tso}{TSO}{transmission system operator}
\newacronym{lib}{LIB}{Li-ion batteries}
\newacronym{lto}{LTO}{Lithium-titanate}
\newacronym{lfp}{LFP}{Lithium iron phosphate}
\newacronym{vfa}{VFA}{value function approximation}
\newacronym{umf}{UMF}{Universal Modelling Framework}

\makenomenclature
\renewcommand\nomgroup[1]{%
  \item[\bfseries
    \ifstrequal{#1}{V}{Physical constants}{%
    \ifstrequal{#1}{N}{Number sets}{%
    \ifstrequal{#1}{A}{Optimization, Control \& Dynamics}{%
    \ifstrequal{#1}{C}{Battery Performance Model}{%
    \ifstrequal{#1}{D}{Battery Degradation Model}{%
    \ifstrequal{#1}{E}{Electric Vehicle (EV) Model}{%
    \ifstrequal{#1}{T}{Thermal Models}
    }}}}}}%
]}

\nomenclature[A]{$w=[w_{\text{grid}}, w_{\text{loss}}, w_{\text{SoC}}, w_{\text{TESS}}]$}{Objective function weights.}
\nomenclature[A]{$\tilde{\ \ }$}{Approximated/estimated variable/function.}
\nomenclature[A]{$\overline{\ \ }, \underline{\ \ }$}{Maximum and minimum variable bounds.}
\nomenclature[A]{$^{+}, ^{-}$}{Positive and negative variable components.}
\nomenclature[A]{$\pi$}{policy}
\nomenclature[A]{$t, t', \Delta t$}{Time index, time index within a policy $\pi$ and timestep.}
\nomenclature[A]{$\mathbb{E}_{W}$}{Expectation with respect to the exogenous process $W$}
\nomenclature[A]{$a \in \mathbb{A}$}{Set of energy assets.}
\nomenclature[A]{$b \in\{\text{BESS}, \text{EV}\} \subset a$ }{Battery storage asset index}
\nomenclature[A]{$X^{\pi}_t(\cdot)$}{Policy function at time $t$}
\nomenclature[A]{$P_{a,t} \in \mathcal{P}$}{Power setpoint for system $a$ at time $t$}
\nomenclature[A]{$\mathcal{P}_{a,[t,t+H]}$}{Power trajectory for system $a$ over horizon $[t,t+H]$}
\nomenclature[A]{$\mathcal{S}_{a,[t,t+H]}$}{State trajectory for system $a$ over horizon $[t,t+H]$}
\nomenclature[A]{$C_{\textrm{grid}}$}{Cost of electricity from the grid}
\nomenclature[A]{$C_{\textrm{loss}}$}{Loss cost or penalty}
\nomenclature[A]{$S_{a,t} \in \mathcal{S}$}{State of system $a$ at time $t$}
\nomenclature[A]{$S^M_{a,t}(\cdot)$}{Transition function for asset $a$ at time $t$}
\nomenclature[A]{$\theta_{a,t}$}{Model parameters for system $a$ at time $t$}
\nomenclature[A]{$W_{t+1}$}{Exogenous information process at time $t+1$}
\nomenclature[A]{$B_{a,t}$}{Belief state for asset $a$ at time $t$}
\nomenclature[A]{$p_{\textrm{SoCDep}}$}{Penalty for EV charging}
\nomenclature[A]{$p_{T_{\textrm{in}}}$}{Penalty for thermal comfort}
\nomenclature[A]{$\varepsilon_{SoC}$}{Deviation from required departure $SoC$}
\nomenclature[A]{$\lambda_{\text{buy/sell},t}$}{Day-ahead buy and sell prices.}
\nomenclature[A]{$T, H$}{Simulation and optimization horizons.}
\nomenclature[A]{$\mathcal{D}_t^{\pi}=[t;t+H]$}{Prediction horizon window at time $t$}

\nomenclature[E]{$\gamma_{\text{EV},t}$}{EV availability at time $t$ (1 = parked, 0 = driving)}
\nomenclature[E]{$t_{\text{dep/arr}} \sim $$\mathcal{T}_{\text{dep/arr}}$}{Departure and arrival time of the EV}
\nomenclature[E]{$P_{\textrm{tot},\textrm{EV},t}$}{Total power associated with EV at time $t$}
\nomenclature[E]{$P_{\textrm{EV},t}$}{EV charger power at time $t$}
\nomenclature[E]{$P_{\text{drive},\text{EV}}$}{Power consumed while driving}
\nomenclature[E]{$SoC_{\textrm{dep}}^*$}{Required SoC at departure}

\nomenclature[C]{$x_{b,t}=i_{b,t}$}{Control cell current to the battery model at time $t$}
\nomenclature[C]{$SoC_{b,t}$}{State of Charge of battery $b$ at time $t$}
\nomenclature[C]{$\tilde{S}^M_{b,t}(\cdot)$}{Battery transition model with performance and ageing}
\nomenclature[C]{$p^M_{b,t}(\cdot)$}{Battery performance model}
\nomenclature[C]{$v_{t,b,t}$}{Terminal voltage of battery $b$}
\nomenclature[C]{$S_{b,t}=[SoC, v_{t}]^T_{b,t}$}{Battery state vector at time $t$ }
\nomenclature[C]{$d^M_{b,t}(\cdot)$}{Battery ageing model}
\nomenclature[C]{$t_{0,b}$}{Battery elapsed lifetime}
\nomenclature[C]{$N_{s,b}$}{Number of cells in series}
\nomenclature[C]{$N_{p,b}$}{Number of parallel branches}
\nomenclature[C]{$\eta_c$}{Coulombic efficiency}
\nomenclature[C]{$Q_{b,t}$}{Battery capacity at time $t$}
\nomenclature[D]{$SoH$}{State of Health of the battery}
\nomenclature[D]{$i_{\text{loss},b,t}$}{Total degradation current}
\nomenclature[D]{$i_{\textrm{SEI},b,t}$}{SEI side reaction current}
\nomenclature[D]{$\eta_{k,b,t}$}{SEI kinetic overpotential}
\nomenclature[C]{$OCV_{b,t}$}{Open-circuit voltage at time $t$}
\nomenclature[D]{$OCV_{n,b,t}$}{Open-circuit voltage of anode}
\nomenclature[D]{$z_{b,t}, z_{b,0\%}, z_{b,100\%}$}{Lithium stoichiometry at time $t$, and limits at 0 and 100\%}
\nomenclature[D]{$i_{\textrm{AM},b,t}$}{Active material loss current}
\nomenclature[D]{$\delta_{\textrm{SEI},b,t}$}{SEI layer thickness}
\nomenclature[D]{$\varepsilon_{e,b,t}$}{Electrolyte volume fraction}

\nomenclature[T]{$T_t$}{Temperature (e.g., $T_{\text{in},t}$, $T_{\text{amb},t}$, $T_{\text{TESS},t}$, $T_{\text{HP},t}$)}
\nomenclature[T]{$\dot{Q}_t$}{Heat transfer rate (e.g., $\dot{Q}_{\text{ir},t}$, $\dot{Q}_{\text{loss},t}$, $\dot{Q}_{\text{cond},t}$, $\dot{Q}_{\text{vent},t}$, $\dot{Q}_{\text{sd},t}$, $\dot{Q}_{\text{cond}}$)}
\nomenclature[T]{$G_{\text{ir},t}$}{Incident solar irradiance at time $t$}
\nomenclature[T]{$s_{T_{\text{in}}}$}{Building temperature slack variable}
\nomenclature[T]{$COP_t$}{Coefficient of performance of the heat pump at time $t$}

\journal{Applied Energy}

\begin{document}

\begin{frontmatter}



\title{Sequential Operation of Residential Energy Hubs using Physics-Based Economic Nonlinear MPC}


\author[inst1]{Darío Slaifstein}
\author[inst1]{Gautham Ram Chandra Mouli}
\author[inst1]{Laura Ramirez-Elizondo}
\author[inst1]{Pavol Bauer}

\affiliation[inst1]{organization={DC Systems, Energy Conversion \& Storage, Electrical Sustainable Energy Department, Delft University of Technology},
            addressline={Mekelweg 8}, 
            city={Delft},
            postcode={2628}, 
            state={Zuid-Holland},
            country={Netherlands}}

\begin{abstract}
The operation of residential energy hubs with multiple energy carriers (electricity, heat, mobility) poses a significant challenge due to different carrier dynamics, hybrid storage coordination and high-dimensional action-spaces. Energy management systems oversee their operation, deciding the set points of the primary control layer. This paper presents a novel 2-stage economic model predictive controller for electrified buildings including physics-based models of the battery degradation and thermal systems. The hierarchical control operates in the Dutch sequential energy markets. In particular common assumptions regarding intra-day markets (auction and continuous-time) are discussed as well as the coupling of the different storage systems. The best control policy it is best to follow continuous time intra-day in the summer and the intra-day auction in the winter. This sequential operation comes at the expense of increased battery degradation. Lastly, under our controller, the realized short-term flexibility of the thermal energy storage is marginal compared to the flexibility delivered by stationary battery pack and electric vehicles with bidirectional charging.
\end{abstract}

\begin{graphicalabstract}
\includegraphics[width=\textwidth]{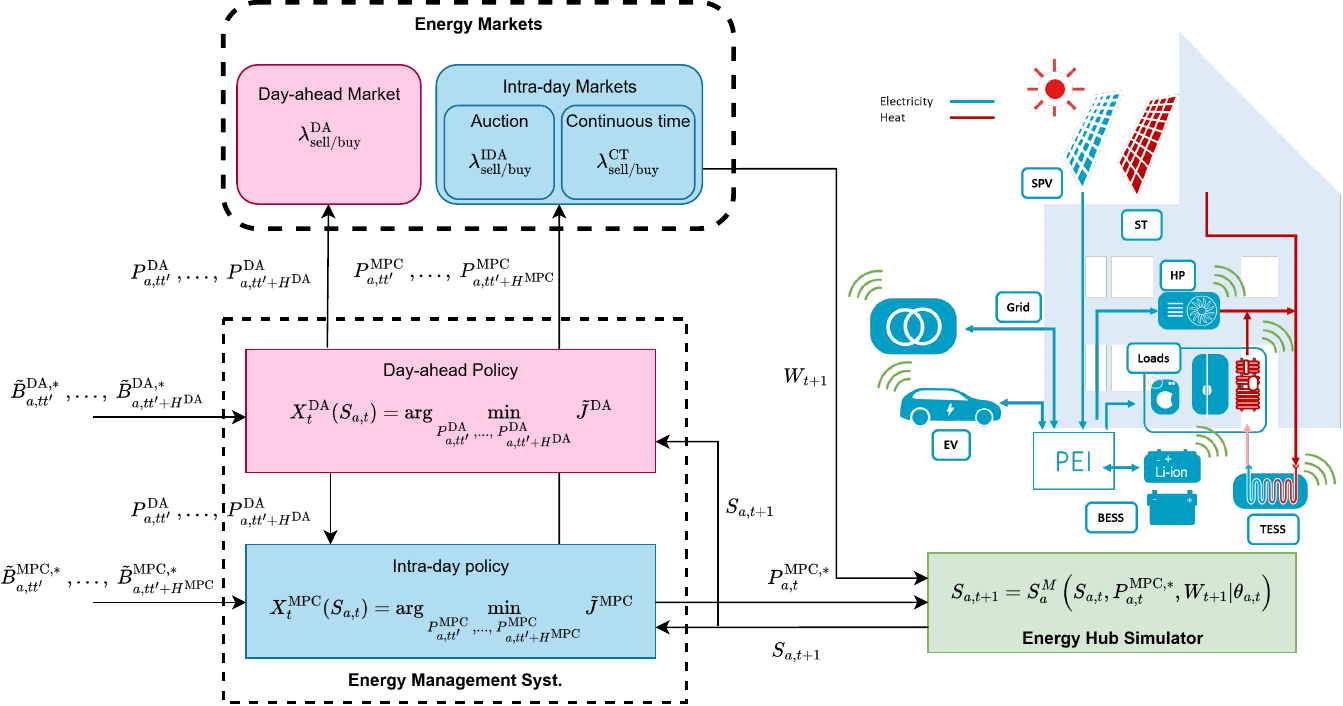}
\end{graphicalabstract}

\begin{highlights}
\item Economic model predictive control using physics-based models ensures building comfort temperature, electric vehicle V2G, and battery ageing control under multiple energy market sequences and flexibility setups.
\item Dutch day-ahead and intra-day markets follow different market dynamics. Strategic operation between auctions and continuous-time, depending on the season, unlocks grid cost savings. The sequential operation increases total battery degradation, challenging common day-ahead assumptions.
\item In a hybrid multi-carrier energy storage system under sequential energy markets, the electrical storage provides the most flexibility. The realized short-term grid value of thermal energy storage is marginal when compared to battery packs and electric vehicles.
\end{highlights}


\end{frontmatter}


\section{Introduction}
\label{sec:intro}
In the context of the energy transition, building electrification poses a significant techno-economic challenge. It is in buildings where different electrification processes intersect most tangibly, with the incorporation of private \ac{ev} and the replacement of traditional gas-boilers with heat-pumps \cite{IEA2024,Li2022}. Harnessing synergies between the different carriers can contribute to more sustainable, flexible, and cost-efficient energy solutions at various levels of the system \cite{Andersson2005, Geidl2007, Vermeer2020, Vermeer2022, Ceusters2021, Ceusters2023, VanDerMeer2018, Ye2020}. To capitalize on such opportunities, system integration and control strategies must be purposefully designed in the \ac{mces}. This integration relies on advanced \ac{ems} capable of coordinating and optimizing the operation of multi-carrier energy storage systems. These must operate within dynamic and uncertain environments in a consistent and reliable manner \cite{EsmaeelNezhad2022, Jouini2024PredictiveStudy}. Moreover, in the future, the participation of these new buildings in energy and power markets appears as an attractive economic opportunity \cite{Jouini2024PredictiveStudy, Garcia-Torres2021OptimalControl, Li2021IntradaySystem}.

\begin{figure*}[tb]
    \centering
    \includegraphics[width=\textwidth]{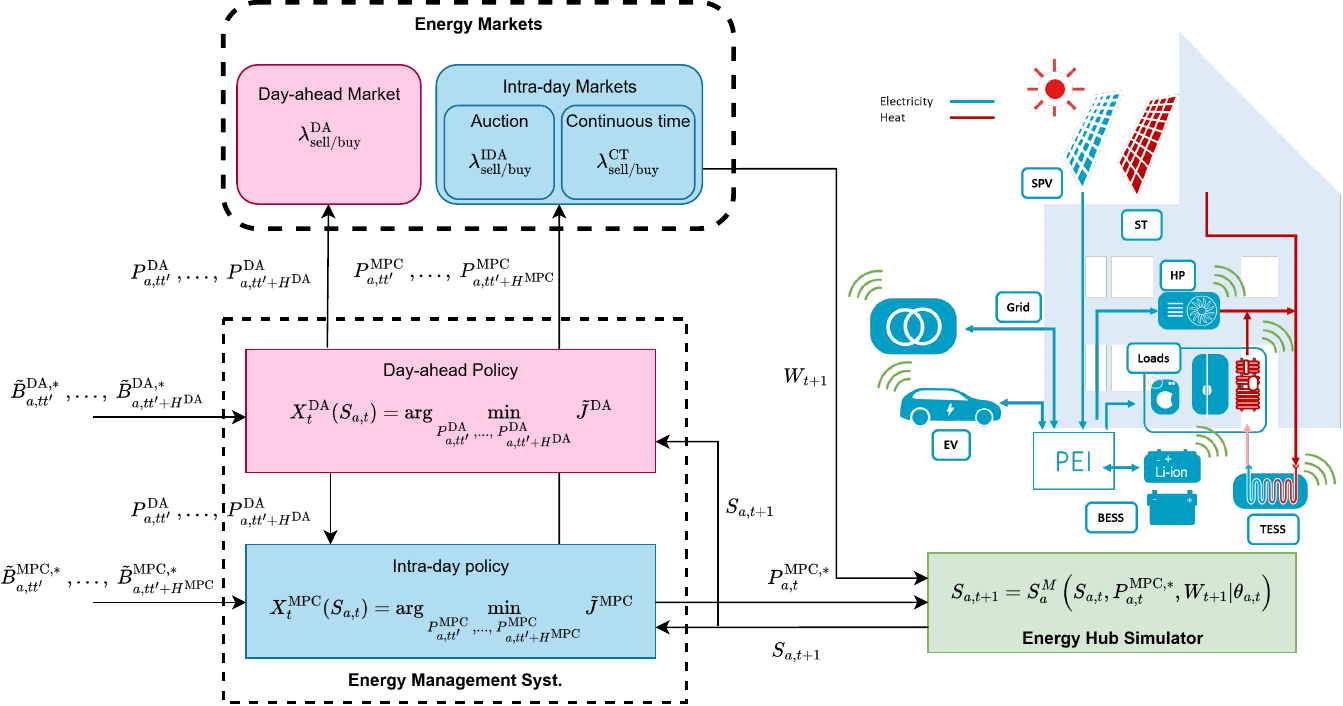}
	\caption{Schematic diagram of the proposed electrified multi-carrier building participating in sequential energy markets.}
	\label{fig:FLXconcept}
\end{figure*}

\section{Literature Review}

 Currently, various sequential electricity markets are implemented across Europe and the US. The \ac{da} market is cleared one day before operation at D-1. Bids are composed of 24hr, 1hr timestep, production and demand schedules. After that, different intra-day markets are opened one or more times (depending on the country) before day D to adjust schedules to recent forecasts. These include pay-as-clear \ac{ida} a couple of hours before the time of delivery and a \ac{ct} market at delivery time with a pay-as-bid mechanism. In auctions, block bids may have different sizes (1-4hrs) and time resolutions (5-15min) depending on the country. The continuous-time intra-day is organized in an order-book which stays open for a day until the time of delivery. Usually, these markets are opened by the \ac{tso} and/or \ac{iso}. On a smaller time scale, different balancing markets are offered by the \ac{tso} and \ac{dso}s. Traditionally, frequency markets are related to \ac{tso}s at the \ac{hv} level, whereas novel local imbalance/congestion markets are being implemented by \ac{dso}s at the MV/LV level. In this paper, the focus is on the day-ahead auction (DA), the intra-day auctions (IDA), and intra-day continuous time (CT). 

Several studies have proposed operating \ac{mces} in energy markets \cite{Jouini2024PredictiveStudy, Li2021IntradaySystem, Zhou2018AMarkets, Johnsen2023TheElectrolyzers, Garcia-Torres2021OptimalControl}. Currently, there are limited options to dispatch and operate residential energy hubs/electrified buildings in the EU or US energy markets due to their current minimum power and energy requirements. Moreover, traditional economic models (marginal costs) have limited capabilities to describe \ac{der} operation in sequential electricity markets. The reader may remember that traditional liberalized energy markets assume non-strategic bidding from their market participants \cite{Schweppe1988SpotElectricity}.  However, it is worth studying the integration of \ac{mces} in sequential energy markets to inform future market designs and stakeholders.

\begin{table*}[bth]
    \centering
    \caption{Summary of Literature Review.}
    \label{tab:litRev}
    \begin{adjustbox}{width=1.1\textwidth}
    \begin{tabular}{l l c c c c c c c c c c l}
        \hline
        & & \multicolumn{3}{c}{\textbf{Loads / Carriers}} & \multicolumn{5}{c}{\textbf{Flexibility}} & \multicolumn{2}{c}{\textbf{Markets}} & \\
        \textbf{Ref.}& \textbf{Application} & \begin{tabular}[c]{@{}l@{}}\textbf{Elec.}\\ \textbf{Load}\end{tabular} & \begin{tabular}[c]{@{}l@{}}\textbf{Space}\\ \textbf{Heating}\end{tabular} & \textbf{Nat.}\textbf{Gas}& \textbf{TESS} & \textbf{EV} & \textbf{BESS} & \begin{tabular}[c]{@{}l@{}}\textbf{Heat}\\ \textbf{Pump}\end{tabular}& \begin{tabular}[c]{@{}l@{}}\textbf{Battery}\\ \textbf{Ageing}\end{tabular}& \textbf{Day-}\textbf{ahead}& \textbf{Intra-}\textbf{day}& \begin{tabular}[c]{@{}l@{}}\textbf{Policy}\\ \textbf{Type}\end{tabular}\\
        \hline
        \cite{Ceusters2021, Ceusters2023, Ceusters2023AnSystems} & \begin{tabular}[c]{@{}l@{}}Multi-Energy\\ Sys.(2MW)\end{tabular} & \cmark & \cmark & \cmark & \cmark & & \cmark & \cmark & & \cmark & & \begin{tabular}[c]{@{}l@{}}Safety\\ focused RL\end{tabular}\\
        \cite{Yu2020} & Building & \cmark & \cmark & & & & \cmark & & & \cmark & & RL-DDPG \\
        \cite{Ahrarinouri2021MultiagentBuildings} & Buildings & \cmark & \cmark & \cmark & & \cmark & & & & \cmark & & MARL \\
        \cite{Huang2022} & Building & \cmark & \cmark & & & & \cmark & & & \cmark & & Safe-MDRL \\
        \cite{Paesschesoone2024ReinforcementUpdates} & \begin{tabular}[c]{@{}l@{}}Industry\\ (4MW)\end{tabular} & \cmark & & & & & \cmark & & & \cmark & & SC-RL \\
        \cite{Zhou2022Data-drivenLearning} & \begin{tabular}[c]{@{}l@{}}Industry\\ (70MW)\end{tabular}& \cmark & \cmark & \cmark & & & \cmark & & & \cmark & & PLSAC-RL \\
        \cite{Jouini2024PredictiveStudy} & \begin{tabular}[c]{@{}l@{}}Multi-Energy\\ (2MW)\end{tabular} & \cmark & \cmark & \cmark & & & \cmark & & & \cmark & \begin{tabular}[c]{@{}l@{}}Scaled\\ prices\end{tabular} & \begin{tabular}[c]{@{}l@{}}2-stage\\ eMPC\end{tabular}\\
        \cite{Li2021IntradaySystem} & \begin{tabular}[c]{@{}l@{}}Multi-Energy\\ (1MW)\end{tabular} & \cmark & \cmark & \cmark & \cmark & & \cmark & \cmark & & \cmark & \begin{tabular}[c]{@{}l@{}}Scaled\\ prices\end{tabular} & \begin{tabular}[c]{@{}l@{}}3-stage\\ hier. MPC\end{tabular} \\
        \cite{Gros2019Day-aheadBuildings} & Building & \cmark & \cmark & \cmark & \cmark & & \cmark & & & \cmark & No trading & \begin{tabular}[c]{@{}l@{}}Schedule\\ \& eMPC\end{tabular} \\
        \cite{Garcia-Torres2021OptimalControl} & \begin{tabular}[c]{@{}l@{}}Microgrid\\ aggregation\end{tabular} & \cmark & & H2 & H2 & & \cmark & & Empirical & \cmark & \begin{tabular}[c]{@{}l@{}}Scaled\\ prices\end{tabular} & \begin{tabular}[c]{@{}l@{}}3-stage\\ eMPC\end{tabular} \\
        \textbf{This work} & Small Building & \cmark & \cmark & & \cmark & \cmark & \cmark & \cmark & \cmark & \cmark & \cmark & \begin{tabular}[c]{@{}l@{}}2-stage\\ eMPC\end{tabular}  \\
        \hline
    \end{tabular}
    \end{adjustbox}
\end{table*}

The main references for this work are presented in Table \ref{tab:litRev}. The literature presents several works dealing with \ac{mces} in sequential markets and ancillary services \cite{Jouini2024PredictiveStudy, Li2021IntradaySystem, Zhou2018AMarkets, Johnsen2023TheElectrolyzers, Garcia-Torres2021OptimalControl}. For day-ahead schedules, Zhou \cite{Zhou2018AMarkets} uses a robust approach to schedule the bids of energy communities in both energy and frequency \ac{da} markets. Similarly, Vermeer \cite{Vermeer2022b} presents a deterministic \ac{da} and FCR for a building with a \ac{bess} and \ac{ev}, without controlling the thermal carrier. Li et al \cite{Li2021IntradaySystem} presents a hierarchical optimization to control an industrial \ac{mces} with power, heat, cooling and gas in 3 different timescales (1hr, 15min and 5min) to capture carrier dynamics, but doesn't include the mobility carrier. Unfortunately, their approach is based only on marginal costs, with no ties to dynamic market prices, and only presents first-order dynamics. References \cite{Ceusters2021,Ceusters2023, Ceusters2023AnSystems, Yu2020, Ahrarinouri2021MultiagentBuildings, Huang2022, Paesschesoone2024ReinforcementUpdates, Zhou2022Data-drivenLearning}, use reinforcement learning techniques to substitute model-based approaches, but the agents only operate in the \ac{da} market and not always include flexible heating systems nor smart \ac{ev} charging. Recently, Jouni \cite{Jouini2024PredictiveStudy} presented a sequential \ac{ems} for \ac{mces} operating in \ac{da} and intra-day energy markets, where the intra-day layer used \ac{empc} \cite{Rawlings2022ModelEdition}. Besides Jouni \cite{Jouini2024PredictiveStudy} the most complete work is by Gros\cite{Gros2019Day-aheadBuildings}. It presents a CHP based microgrid and a 2-stage \ac{mpc} to control it. Unfortunately, the heating is not electrified, and its not integrated to intra-day markets. All references that take into account intra-day markets refer only to the continuous-time (CT) intra-day market, assuming that this market follows the same dynamics as the day-ahead market and only scales the day-ahead prices. This assumption disregards the different frequency components and volatility of the \ac{da}, \ac{ida} and \ac{ct} leading to a poor dynamic analysis of the storage systems used. Finally, only 1 (one) work accounts for \ac{ev} integration \cite{Ahrarinouri2021MultiagentBuildings} but disregards numerous other aspects (intra-day market, \ac{hess}, battery ageing, etc.). The interaction between the different storage systems (\ac{bess}, \ac{ev}, and \ac{tess}) depends on the various price signals being followed. Hence, how these elements interact is crucial to understanding multi-carrier systems.

Regarding the current white-box or \ac{mpc} policy approaches, their dynamic models are limited to simplified linear time-invariant systems. Moreover, most works only focus on one carrier at a time \cite{Vermeer2020, Vermeer2022b, Li2019, Risbeck2018Mixed-IntegerSystems}. To coordinate \ac{hess}, the different technologies must be modeled, representing power limits, dynamics, and other particularities. Heating models focusing on thermal comfort are based on building and device thermodynamics \cite{Mariano-Hernandez2021ADiagnosis, Risbeck2018Mixed-IntegerSystems, Damianakis2023Risk-averseConsumption, Alpizar-Castillo2024ModellingHouse}. A \ac{tess} will have different dynamics depending on its technology and a \ac{hp} will present nonlinear conversion efficiencies if they are air-air, air-water, and so on \cite{Alpizar-Castillo2024ModellingHouse, Damianakis2023Risk-averseConsumption}. To model the heating approximating it as a flow can be an option. Nevertheless, it assumes a fixed temperature setpoint, reducing operational flexibility and hinders the tractability of the underlying optimal control problem \cite{Ceusters2023, Zhou2022Data-drivenLearning, Jouini2024PredictiveStudy, Li2021IntradaySystem}.

Another critical modelling point is battery degradation, as its operation affects the lifetime; thus, ambitious controls might hinder it. Detailed battery models are usually reserved for local controls, whereas \ac{ems} formulations for residential \ac{mces} tend to simplify the models to linear or quadratic forms, overlooking most technology particularities like chemistry, diffusion dynamics, or capacity fade \cite{Geidl2007, Ye2020, Ceusters2021, Mariano-Hernandez2021ADiagnosis, Damianakis2024CoordinatedGrids}.  Implementing approximated empirical models that are not meant for control applications is the standard practice \cite{VegaGarita2025TheReview}. Unfortunately, such degradation models only have interpolation capabilities, typically use non-linear equations, represent a limited number of operating conditions (average C-rate, minimum $SoC$, etc.), are prone to overfitting, and are chemistry dependent. On the other hand, physics-based (PB) models are built through first-principles and specialized tests to identify individual degradation mechanisms \cite{Okane2022, Prada2013Simulations, Reniers2019, Plett2024BatteryMethods}. They have extrapolation features, can be expressed in the state-space form,  account for several cathode chemistries, and represent a wide range of operating conditions. Even though they are non-linear and non-convex, they have been integrated into different optimal control schemes through control-oriented \ac{pbrom} and demonstrating bigger potential cost savings than their empirical counterparts \cite{Okane2022, Jin2017, Purewal2014DegradationModel, Reniers2019, Prada2013Simulations, Xavier2021, Reniers2023, Reniers2021, Jin2022, Li2019, Dorronsoro2025BatteryModels, Slaifstein2025Aging-awareSystems, VegaGarita2025TheReview}. Overlooking these modelling assumptions (battery degradation and thermal controls) hinders the accuracy, generality, and practicality of the results in the literature, since potential interactions between systems are limited by them.

Summing up, five main gaps can be identified in the literature:
\begin{enumerate}
    \item Usually, intra-day markets are only addressed by scaling the prices of the \ac{da} market. This simplification does not consider the differences between intra-day auctions and continuous time intra-day, especially in their frequency spectrum, leading to sub-optimal decisions \cite{Hornek2025TheStrategies, Birkeland2024QuantifyingMarket}. Is it still true for integrated energy systems? How does it impact battery degradation?
    \item Integration of electrified buildings with integrated mobility in sequential markets has been introduced,  but it has not included bidirectional charging, nor its stochastic availability. Improving on these aspects presents an opportunity for additional flexibility and added value.
    \item  Joint operation of hybrid energy storage systems that combine \ac{bess}, \ac{tess}, and \ac{ev} is uncommon. Integrated operation could unlock synergies between carriers and storage systems. In other words, does individual \ac{ess} performance prevail, or is there a limitation in their interaction?
    \item Battery degradation has been studied using empirical models that were not meant for dynamic operation and integration in \ac{ems} schemes. This leads to suboptimal results and reduced flexibility \cite{VegaGarita2025TheReview}. For example, preventing battery degradation through \ac{sei} control incentivizes the battery to be discharged $SoC \rightarrow \underline{SoC}$ \cite{Plett2024BatteryMethods, Slaifstein2025Aging-awareSystems, Xavier2021}. Does this affect the rest of the \ac{hess}? Is this still valid under higher frequency prices? Can degradation be controlled without sacrificing grid benefits \cite{Reniers2021}?
    \item Detailed thermal modelling is often reserved for studies where single-carrier systems are analyzed. Accounting for nonlinear \ac{hp} behavior and utilizing a comfort band to control temperature unlocks potential savings and flexibilities that actively interact with \ac{hess} components.
\end{enumerate}

To address such gaps the contributions of this paper are:
\begin{enumerate}
    \item   A novel two-level economic model predictive control \ac{ems} for residential energy hubs that integrates: day-ahead and intra-day markets, \ac{pbrom} ageing models for battery-based \ac{ess}, flexible electrical heating control, and \ac{ev} bidirectional smart-charging.
    \item An in-depth analysis of the residential energy hub participation in the day-ahead and intra-day markets using real data from \cite{2023EPEXServices}.
    \item  A detailed analysis of the interaction between electric and thermal carriers, including \ac{bess}, \ac{ev}, and \ac{tess}, in the context of electrified buildings under deterministic and noisy settings.
    \item An open-source library for implementing optimization-based \ac{ems} for multiple applications.
\end{enumerate}

A schematic of the system under study is presented in Fig.~\ref{fig:FLXconcept}. The building is composed of solar photovoltaics (SPV), a battery energy storage system (BESS), an electric vehicle (EV), a power electronic interface (PEI), a heat pump (HP), a thermal energy storage system (TESS), a grid connection, and loads. This paper focuses on optimizing the flows behind this grid connection for a single system, with the aim to show the potential behind such applications. Each level of the \ac{ems} corresponds to an energy market. The first layer consists of a planner participating in the day-ahead market, and the second layer is an \ac{empc} participating in the intra-day markets. 

This manuscript is organized as follows: section \ref{sec:modelling} presents the problem and modelling framework, section \ref{sec:policyDLA} presents the algorithm design and models used; section \ref{sec:results} describes our case studies and validation; finally section \ref{sec:conclusions} presents the conclusions and future works.

\section{Sequential Market Models}
\label{sec:modelling}

\begin{figure}
    \centering
    \includegraphics[width=\linewidth]{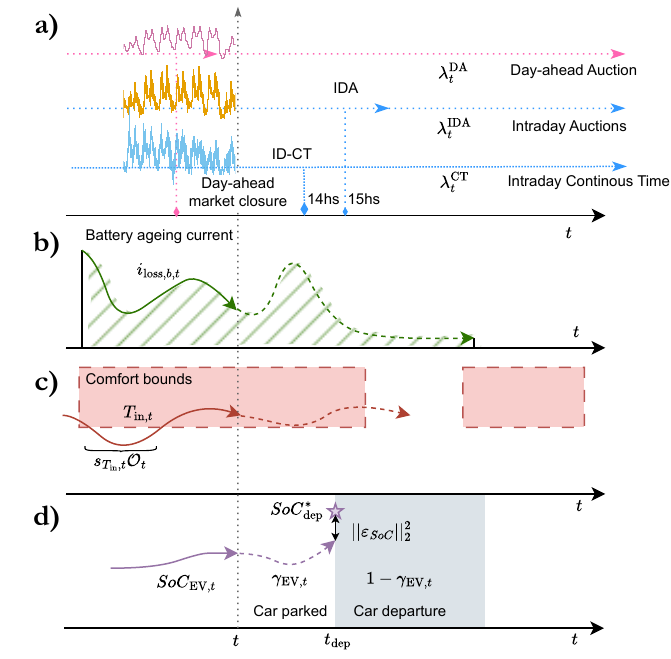}
    \caption{Components of the cost function: a) grid cost $C_{\text{grid}}$, b) battery degradation cost $C_{\text{loss}}$, c) thermal comfort penalty $p_{T}$ and d) electric vehicle charging penalty $p_{\text{SoCDep}}$.}
    \label{fig:costFunction}
\end{figure}

 The following section describes the \ac{ems} models, following the \ac{umf} by Powell \cite{Powell2019, Powell2022} and the models developed in \cite{Slaifstein2025Aging-awareSystems}. For a given system size, the objective is to handle the operation cost, which is composed of four parts: the net cost of energy from the grid $C_{\textrm{grid}}$, the degradation cost of losing storage capacity $C_{\textrm{loss}}$, a penalty for not charging the \ac{ev} at departure $p_{\textrm{SoCDep}}$ and a penalty to ensure thermal comfort $p_T$. The grid cost, the degradation cost and the thermal discomfort are cumulative objectives because the goal is to optimize them through time, while the penalty for not charging the \ac{ev} to the desired $SoC$ is only a point penalty at departure times $t_{\textrm{dep}}$. The \ac{sdp} is then:
\begin{subequations}\label{eq:sdp}
    \begin{align}
        \min_{x^*_{a,t}} \quad & \mathbb{E}_{W}[ C_{\textrm{grid}}+C_{\textrm{loss}}+p_{\textrm{SoCDep}}+p_T]\\
        \textrm{s.t.} \quad & S_{a,t+1}=S^M_{a,t} (S_{a,t}, x^*_{a,t}, W_{t+1}|\theta_{a,t}) \\
                    & x_{a,t}^*=X^{\pi}_t(S_{a,t}) \in \mathcal{X} & \forall a \in \mathbb{A},\ t \in \mathcal{D}_t \\
                    & S_{a,t} \in \mathcal{S} & \forall a \in \mathbb{A},\ t \in \mathcal{D}_t \\
                    & \mathbb{A}=\{\text{SPV}, \text{grid}, \text{EV}, \text{BESS}, \text{HP}, \text{TESS}\}
    \end{align}
\end{subequations}

\noindent{where $S_{a,t}$ is the state vector, $x^*_{a,t}$ is the optimal decision for timestep $t$, $W_{t+1}$ is an exogenous process that introduces new information after making a decision. The mappings $S_{a,t}^M(\cdot)$, and $X^{\pi}_t(\cdot)$ are the transition function and optimal policy, respectively. The first is a set of equations describing the states and parameter evolution, and the second is the algorithm that finds the setpoints. The vector $\theta_{a,t}$ contains all the parameters of each asset $a$ and changes over time $t$. The subindex $a \in \mathbb{A}$ corresponds to the assets shown in Fig. \ref{fig:FLXconcept}. The index $b \in\{\text{BESS}, \text{EV}\} \subset a$ denotes the electric storage assets.  The simulation time is $T$ and the timestep $\Delta t$ and the time domain is $\mathcal{D}_t=[0,\Delta t,...,T]$. }.

The components of the objective are presented in Figure \ref{fig:costFunction}. From top to bottom the first component is grid cost $C_{\text{grid}}$ defined by the cost in each market:
\begin{subequations}
\begin{align}
\label{eq:cgrid}
    & C_{\textrm{grid}} = C_{\textrm{grid}}^{\textrm{DA}} + C_{\textrm{grid}}^{\textrm{MPC}}\\
& C_{\textrm{grid}}^{\textrm{DA/MPC}}=w_{\text{grid}}\sum_{t=0}^{T}c_{\textrm{grid},t}^{\textrm{DA/MPC}}.\Delta t \\
 \label{eq:cgridDA}
& c_{\textrm{grid},t}^{\textrm{DA}} = \frac{\lambda_{\textrm{buy},t}^{\textrm{DA}} - \lambda_{\textrm{sell},t}^{\textrm{DA}}}{2} |P_{\textrm{grid},t}^{\textrm{DA}}|+ \frac{\lambda_{\textrm{buy},t}^{\textrm{DA}} + \lambda_{\textrm{sell},t}^{\textrm{DA}}}{2} P_{\textrm{grid},t}^{\textrm{DA}} \\
\label{eq:cgridCT}
    & c_{\textrm{grid},t}^{\textrm{MPC}} =  \frac{\lambda_{\textrm{buy},t}^{\textrm{MPC}} - \lambda_{\textrm{sell},t}^{\textrm{MPC}}}{2} \left| \Delta P_{\textrm{grid},t} \right|+ \frac{\lambda_{\textrm{buy},t}^{\textrm{MPC}} + \lambda_{\textrm{sell},t}^{\textrm{MPC}}}{2}  \Delta P_{\textrm{grid},t} \\
     \label{eq:deltaPg}
     &  \Delta P_{\textrm{grid},t} = P_{\textrm{grid},t}^{\textrm{MPC}} - P_{\textrm{grid},t}^{\textrm{DA}}
\end{align}
\end{subequations}

The superscripts \ac{da} and \ac{mpc} denote the layer of the controller, with the \ac{mpc} layer connected to an intra-day market. The subscripts "buy" and "sell" denote each price, $w_{\text{grid}}$ is a user-defined weight, $\lambda_t$ are the market prices, and $P_{\text{grid},t}$ is the grid power. The formulation from \cite{Jouini2024PredictiveStudy, Garcia-Torres2021OptimalControl} is chosen due to its robustness and numerical properties. In this work, only one \ac{ida} is used, and the \ac{ct} order book is approximated through the ID1 index, which is the average of all transactions in the order book for the last hour.

The second component is the battery degradation cost $C_{\text{loss}}$. For each battery $b$ a capacity fade current $i_{\text{loss},b,t}$ represents how many Ah are lost per unit time per cell. Assuming that all cells are perfectly balanced, each pack has $N_{s,b}$ cells in series per branch and $N_{p,b}$ branches in parallel, the total degradation is defined as:
\begin{equation}
        C_{\text{loss}}=w_{\text{loss}}.c_{\text{loss}}.\sum_{t=0}^{T}\sum_{b}{N_{\textrm{s},b} N_{\textrm{p},b}i_{\text{loss},b,t}.\Delta t},\ \forall\ b \subset a,
\end{equation}
\noindent{the marginal cost is $c_{\text{loss}}=1.2$ \texteuro/Ah \cite{Reniers2023}, and a weight $w_{\text{loss}}$ for tuning, scaling and blending all components together.}

The third objective is the user's thermal comfort. The building's internal temperature $T_{\text{in},t}$ has to be kept within bounds $\underline{T}_{\text{in}}, \overline{T}_{\text{in}}$ while the building is occupied. Any excursion outside of these bounds should be penalized. A slack variable $s_{T_{in},t}$ is defined to measure temperature excursions and the goal is to minimize it over time as in:
\begin{subequations}
    \begin{align}
    & s_{T_{in},t}=\max\left(0, \max \left( \underline{T}_{\text{in}} - T_{\text{in},t}, T_{\text{in},t} - \overline{T}_{\text{in}} \right) \right)\\
        & p_T  = w_T \sum_{t=0}^{T} s_{T_{\text{in}},t}. \mathcal{O}_t . \Delta t,\
    \end{align}
    \label{eq:thrmComf}
\end{subequations}

\noindent{the building occupancy is $\mathcal{O}_t$ defined as 1 for $t$ when there's someone inside the building and 0 when nobody is.}

Lastly, the \ac{ev} must have enough energy to drive while it's away from home. The user defines a setpoint for departure $SoC_{\text{dep}}^*$ and the \ac{ev} has to reach it at departure time $t_{\text{dep}}$. The measure to be minimized is then the distance between $SoC_{\text{EV},t_{\text{dep}}}$ and $SoC_{\text{dep}}^*$, weighted by $w_{SoC}$:
\begin{subequations}\label{eq:evpen}
    \begin{align}
    & \varepsilon_{SoC}=SoC_{\text{EV},t_{\text{dep}}}-SoC_{\text{dep}}^* \\
    & p_{\textrm{SoCDep}}=w_{SoC}.||\varepsilon_{SoC}||_2^2,
    \end{align}
\end{subequations}

Following the definitions in \cite{Slaifstein2025Aging-awareSystems} the state vector has physical measurements $R_{a,t}$ and beliefs $\tilde{B}_{a,t}$ that approximate the exogenous process $W_{t+1}$ as $S_{a,t}=[R_{a},\tilde{B}_{a}]_t^T $, with $\tilde{B}_{a,t}=[\tilde{G}_{\text{ir}},  \tilde{\gamma}_{\text{EV}}, \tilde{P}_{\text{load}}, \tilde{\mathcal{O}}, \tilde{T}_{\text{amb}}]_t^T$.  The actions or decision variables are $x^*_{a,t}=[P_{\text{EV}}, P_{\text{BESS}}, P_{\text{HP}}, \dot{Q}_{\textrm{HP}}^{\text{D}},\dot{Q}_{\textrm{TESS}}^{\textrm{D}}]_t^T$. Both the actions and state vectors have upper and lower limits denoted as $\overline{x}_{a,t}^*$, $\underline{x}_{a,t}^*$, $\overline{S}_{a,t}$, and $\underline{S}_{a,t}$. To account for converter efficiencies $\eta_a$, bidirectional powers, either actions or states, are modeled as $P_{a,t} = \eta_a P^+_{a,t} - \frac{1}{\eta_a} P^-_{a,t}$, and complementarity constraints  $P_{a}^+ \perp P_{a}^-$.

\section{Policy design}\label{sec:policyDLA}

\begin{figure}[tb!]
    \centering
    \includegraphics[width=\columnwidth]{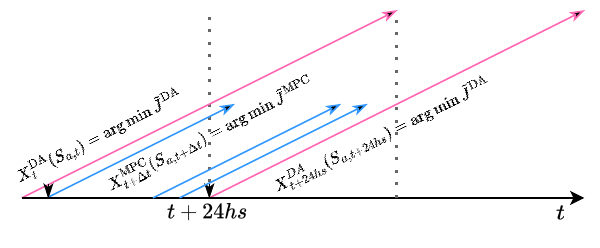}
    \caption{Deterministic DLA seq. policy with a day-ahead planner and an MPC.}
    \label{fig:appDLA}
\end{figure}

\begin{algorithm}[b!]
    \caption{Sequential market operation algorithm}\label{alg:seqMkt}
    \begin{algorithmic}[1]
	\State \textbf{Initialize hyperparameters} $\Delta t,\ H^{\text{DA/MPC}},\ w,\ n_d$ \;
    \State \textbf{Initialize device states and inputs} $S_{a,\ 0}$ \;
    \For{$d \in 1:n_d$}
        \State Solve the deterministic OCP-DA, Eq. \ref{eq:ocpDA}, and obtain schedule $\mathcal{P}_{a,[t,H^{\text{DA}}]}^{\textrm{DA}}$. \;
        \For{$t \in 0:H^{\text{MPC}}$}
        \State Solve the deterministic OCP-CT, Eq. \ref{eq:ocpCT}, and obtain action $P_{a,t}^{\textrm{MPC}}$. 
        \State Simulate $S_{a,t+1}=S_{a,t}^M(S_{a,t}, P^{\textrm{MPC}}_{a,t}, W_{t+1})\ $;
        \State Update forecasts in $B_{a,tt'}^{\text{MPC}}$\;
        \State Move time window $t \leftarrow t + \Delta t^{\text{MPC}}$;\;
        \EndFor
    \EndFor
    \end{algorithmic}
\end{algorithm}

The \ac{sdp} in Eq.\ref{eq:sdp} is a state-dependent problem where current states influence future decisions. To solve such \ac{sdp} a hierarchical policy $\pi$ is proposed. The process is shown in Fig. \ref{fig:appDLA} and explained in Algorithm \ref{alg:seqMkt}. Two policies are sequentially applied, first a day-ahead policy $X_t^{\textrm{DA}}$ offers a 24hr schedule for the day-ahead market $\mathcal{P}_{a,[t,t+24]}^{\textrm{DA}}$, after $\Delta t$ an \ac{empc} policy $X_{t+\Delta t}^{\textrm{MPC}}$ updates the \ac{da} setpoint with the new information (forecasts, states, etc.), placing a bid in the intra-day market. This updated setpoint is implemented in the real system or simulator $S^{M}_{a,t}$. This is repeated every timestep until the moment of presenting a new \ac{da} schedule is reached at $t+24$hr.

The two policies are based on approximated state-space models $\tilde{S}_{a,t}^{M}$ of the \ac{mces}. First, the \ac{da} policy optimizes actions $P_{a,tt'}^{\textrm{DA}}$ over the lookahead time $t'$, with horizon $H^{\textrm{DA}}$ and timestep $\Delta t^{\text{DA}}$:
\begin{subequations}\label{eq:ocpDA}
    \begin{align}
        \min_{P^{\textrm{DA}}_{a,tt'}} \quad & \tilde{J}^{\textrm{DA}} \\
        \textrm{s.t.} \quad & \tilde{S}_{a,tt'+1}^{\textrm{DA}}=\tilde{S}^{M}_a \left( \tilde{S}_{a,tt'}^{\textrm{DA}}, P^{\textrm{DA}}_{a,tt'}, \tilde{B}^{\textrm{DA}}_{tt'}|\theta_{a,tt'} \right) \\
                    & \tilde{SoC}_{\textrm{BESS},tt'_1}^{\textrm{DA}} = \tilde{SoC}_{\textrm{BESS},tt'_1+24hs}^{\textrm{DA}} \label{eq:pcDA} 
    \end{align}
\end{subequations}

Later, the \ac{empc} optimizes actions $P_{a,tt'}^{\textrm{MPC}}$ over the lookahead time $t'$, with horizon $H^{\textrm{MPC}}$ and timestep $\Delta t^{\text{MPC}}$:
\begin{subequations}\label{eq:ocpCT}
    \begin{align}
        \min_{P^{\textrm{MPC}}_{a,t}} \quad & \tilde{J}^{\textrm{MPC}}\\
        \textrm{s.t.} \quad & \tilde{S}_{a,tt'+1}^{\textrm{MPC}}=\tilde{S}^{M}_a \left( \tilde{S}_{a,tt'}^{\textrm{MPC}}, P^{\textrm{MPC}}_{a,t}, \tilde{B}^{\textrm{MPC}}_{tt'}|\theta_{a,tt'}\right) \\
                    & \tilde{SoC}_{\textrm{BESS},tt'_0}^{\textrm{MPC}} = \tilde{SoC}_{\textrm{BESS},tt'_0+H^{\textrm{MPC}}}^{\textrm{MPC}} \label{eq:pcCT} 
    \end{align}
\end{subequations}

Where both objective functions are:
\begin{equation}
    \tilde{J}^{\textrm{DA/MPC}} = \tilde{C}_{\textrm{grid}}^{\textrm{DA/MPC}}+ \tilde{C}_{\textrm{loss}}^{\textrm{DA/MPC}}+\tilde{p}_{\textrm{SoCDep}}^{\textrm{DA/MPC}}+\tilde{p}_T^{\textrm{DA/MPC}}
\end{equation}

\noindent{where the tilde $\tilde{}$ denotes approximate, the time $t$ is the time at which the \ac{dla} policy is created and $t'$ is the time inside the policy itself and superscripts \ac{da} and \ac{mpc} mark to which policy the variables correspond to. The main differences between $X_{a,t}^{\textrm{{DA}}}$ and $X_{a,t}^{\textrm{{MPC}}}$ are: their sampling frequency $\Delta t^{\textrm{DA}}=1$hr and  $\Delta t^{\textrm{MPC}}=15$min, their prediction horizon  $H^{\textrm{DA}}=48$hr and  $H^{\textrm{MPC}}=24$hr, their grid cost functions $C_{\textrm{grid}}^{\textrm{DA/MPC}}$, Eqs. \ref{eq:cgridDA} and \ref{eq:cgridCT}, and their periodicity conditions, Eqs. \ref{eq:pcDA} and \ref{eq:pcCT}. These conditions mean that in the day-ahead policy $X^{\text{DA}}_t$ solved at time $t$, within the lookahead time $t'$ the  $\tilde{SoC}_{\text{BESS}}^{\textrm{DA}}$ at policy time $t'_1$ has to be the same at time $t'_1+24$hr. Eq. \ref{eq:pcCT} means that the estimated \ac{bess} state $\tilde{SoC}_{\text{BESS}}^{\textrm{MPC}}$ at the initial policy time $t'_0$ has to be equal to the final state at the end of the horizon $t'_0+H^{\text{MPC}}$. These two constraints are key for bounding the corresponding value functions and ensuring their bounds \cite{Grune2017NonlinearAlgorithms}. Combined with the optimization horizons $H^{\text{DA/MPC}}$ these ensures the turnpike properties of the controller.}

The sequential deterministic optimizations approximate the true \ac{sdp} in Eq. \ref{eq:sdp} by using forecasts, stored in $\tilde{B}_{a,tt'}^{\textrm{DA/MPC}}$, and approximated models for the transition functions $\tilde{S}_{a,t}^{M}$. The approximate transition function $\tilde{S}_{a,t}^M(.)$ is the compendium of the equations specified in the following sections. The controller's actions are evaluated in the true transition function $S_{a,t}^M$, defined in \cite{Slaifstein2025Aging-awareSystems}. The simulator enforces all the thermal dynamics presented in Section \ref{sse:thermal} \cite{Alpizar-Castillo2024ModellingHouse} and the battery dynamics through high-fidelity models \cite{Planden2022}. Note the subtle difference between the approximated dynamics $\tilde{S}_{a,t}^M$ and the real ones $S_{a,t}^M$. This is not to be overlooked because the assumption that the predictions made by the policy $\pi$ hold true can lead to disappointing results in real-world applications. In the future, the simulator might grow enough to be considered a digital twin of the real building.

Thus, the policies are:
\begin{equation}
    X^{\textrm{DA/MPC}}_t(S_{a,t})= \arg \min_{P_{a,t}^{\textrm{DA/MPC}}}  \tilde{J}^{\textrm{DA/MPC}}
\end{equation}
 In the remainder of this section, all variables will be presented without approximates  " $\tilde{}$ "  or layer superscripts DA or MPC, since all the models are present in both the \ac{ems} policies $X^{\textrm{DA/MPC}}_t$ and the simulator $S^{M}_{a,t}$.

\subsection{Thermal carrier}\label{sse:thermal}
\subsubsection{Building}\label{ssse:building}

\begin{figure}[b!]
    \centering
    \includegraphics[width=0.9\columnwidth]{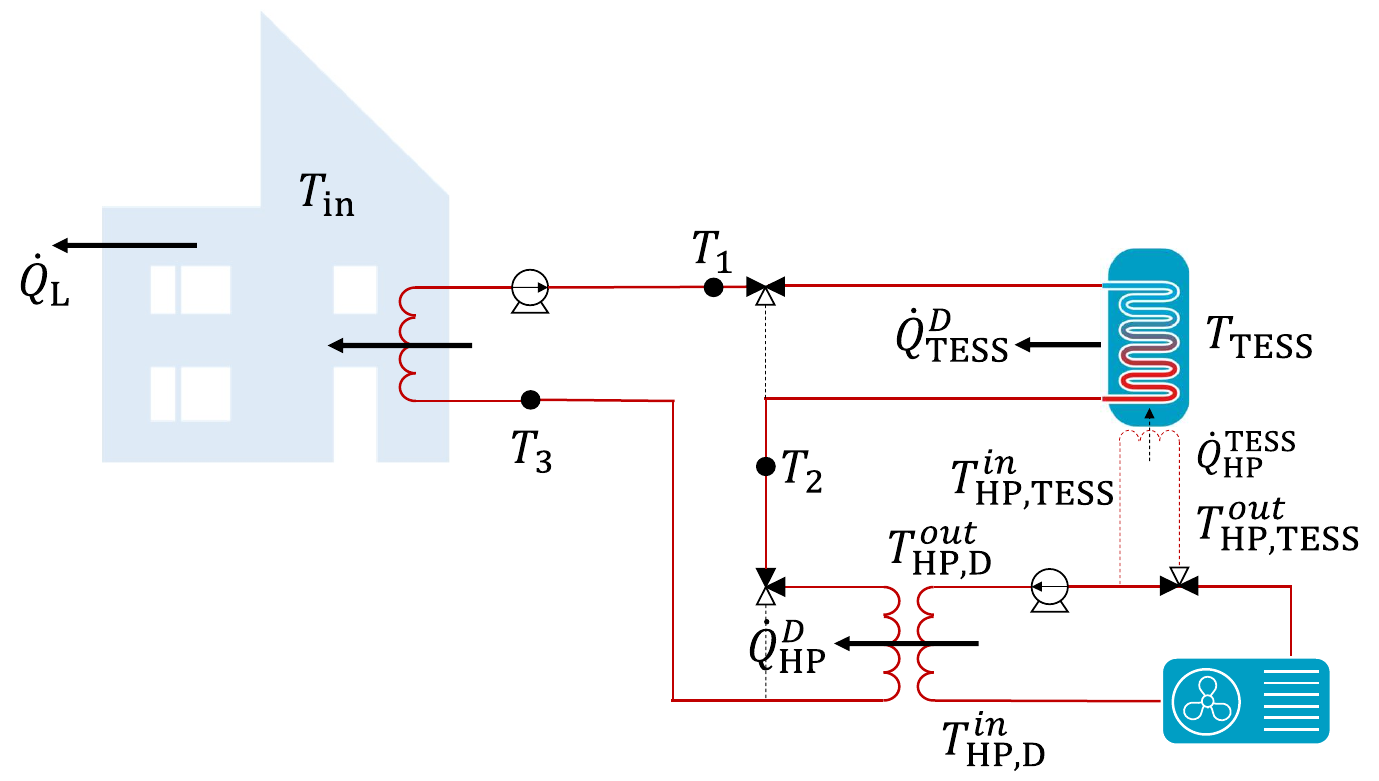}
    \caption{Thermal heating system.}
    \label{fig:thermal}
\end{figure}

The building has an electrical heating system, presented in Fig. \ref{fig:thermal} . The system comprises a \ac{hp} to generate heat, a \ac{tess} to store it, and radiators to distribute it. The building loses heat $\dot{Q}_{\textrm{loss},t}$ through its ventilation $\dot{Q}_{\textrm{vent},t}$ and conduction $\dot{Q}_{\textrm{cond},t}$ losses. Temperature/potential-based models are used to design the controls. Another alternative are power/flow-based models such as the ones used in \cite{SlaifsteinIECON23, Slaifstein2025Aging-awareSystems}. In the case of the latter, the problem becomes a scheduling problem (supply-demand matching), whereas the first option sets up a soft-tracking objective in which the building's inner temperature $T_{\text{in},t}$ is maintained within bounds, Eq. \ref{eq:thrmComf}. The soft-tracking problem defined by temperature-based models is less computationally complex, as the defined terminal set and corresponding value function are bounded \cite{Grune2017NonlinearAlgorithms}. This design choice simplifies the computational complexity of the policies $X^{\textrm{DA/MPC}}_t(\cdot)$.

The building's thermal model dictates the evolution of its internal temperature $T_{\text{in},t}$ following a linear state-space model:
\begin{subequations}\label{eq:TinDyn}
\begin{align}
    & S_{\text{in},t}=T_{\text{in},t},\ x_{\text{in},t}=[\dot{Q}_{\text{TESS}}^{\text{D}},\ \dot{Q}_{\text{HP}}^{\text{D}}]^T_t,\ W_{\text{in},t+1}=[G_{\text{ir}},\ T_{\text{amb}}]^T_t \\
    & S_{\text{in},t+1}=\textbf{A}_{\text{in}} S_{\text{in},t}+\textbf{B}_{\text{in}}x_{\text{in},t}+\textbf{D}_{\text{in}}W_{\text{in},t+1}
\end{align}
\end{subequations}

\noindent{where $T_{\text{amb},t}$ is the ambient temperature,  $G_{\textrm{ir},t}$ is the global irradiance, $\dot{Q}_{\textrm{loss},t}$ are the losses, $\dot{Q}_{\textrm{TESS},t}^{\textrm{D}}$ is the heat supplied by the \ac{tess} to the building,  $\dot{Q}_{\textrm{HP},t}^{\textrm{D}}$ is the heat supplied by the \ac{hp} to the building. The losses are assumed to be linear with the temperature difference \cite{Damianakis2023Risk-averseConsumption}. All temperatures are expressed in [K], and all heat flows are in [kW]. In the model Eq. \ref{eq:TinDyn}, the exogenous information $W_{\text{in},t+1}$ is the ambient temperature $T_{\text{amb},t}$ and global irradiance $G_{\textrm{ir},t}$. The actions $x_{\text{in},t}$ are the heat flows $\dot{Q}_{\textrm{HP},t}^{\textrm{D}}$ and $\dot{Q}_{\textrm{TESS},t}^{\textrm{D}}$. Losses and incident radiation heat follow the models in \cite{Damianakis2023Risk-averseConsumption, Alpizar-Castillo2024ModellingHouse}, and are summarized in \ref{sec:appA}. }

\subsubsection{Heat Pump}

An air-to-water \ac{hp} generates heat from electrical power following a non-linear model \cite{Damianakis2023Risk-averseConsumption}. The model considers complementarity between heating the \ac{tess} or the building with distinct temperatures and a non-linear coefficient of performance. The model is summarized as:
\begin{subequations}\label{eq:hp}
    \begin{align}
    & \dot{Q}_{\text{HP},t} = COP_{t}.P_{\text{HP},t}\,, \\
    & COP_t = 7.90471 . e^{-0.024 . (T_{\textrm{HP},t}^{\text{in}} - T_{\text{amb},t})} \\
    & \dot{Q}_{\text{HP},t} = \dot{Q}_{\text{HP},t}^{\textrm{TESS}} + \dot{Q}_{\text{HP},t}^{\textrm{D}} \\
    & \dot{Q}_{\text{HP},t}^{\textrm{TESS}} \perp \dot{Q}_{\text{HP},t}^{\textrm{D}} \\
    & T_{\text{HP,D},t}^{\textrm{in}} = T_{\text{HP,D},t}^{\textrm{out}} - \frac{\dot{Q}_{\text{HP}}^{\text{D}}}{\eta_{\text{HP}} . \dot{m}_f . c_f} \\
    & T_{\text{HP, TESS},t}^{\textrm{in}} = T_{\text{HP, TESS},t}^{\textrm{out}} - \frac{\dot{Q}_{\text{HP}}^{\text{TESS}}}{\eta_{\text{HP}} . \dot{m}_f . c_f} \\
    & T_{\text{HP},t}^{\textrm{in}}=\begin{cases}
      T_{\text{HP, D},t}^{\textrm{in}} & \dot{Q}_{\text{HP},t}^{\textrm{D}} \neq 0 \\
      T_{\text{HP, TESS},t}^{\textrm{in}} &  \dot{Q}_{\text{HP},t}^{\textrm{TESS}} \neq 0
    \end{cases} \,.
    \end{align}
\end{subequations}

\noindent{where $\dot{Q}_{\textrm{HP}}$ is the total heat produced by the \ac{hp}, $COP_t$ is the coefficient of performance,  $P_{\textrm{HP},t}$ is the consumed electrical power, $T_{\textrm{HP},t}^{\text{in/out}}$ is the inlet/outlet temperature of the \ac{hp},  $\dot{Q}_{\text{HP},t}^{\textrm{D}}$ is the heat supplied to the building, and $\dot{Q}_{\text{HP},t}^{\textrm{TESS}}$ is the heat supplied to the \ac{tess}. These last two are complementary with independent heat exchangers parallel to each other. The non-linear $COP_t$ model, from \cite{Damianakis2023Risk-averseConsumption}, uses $T_{\textrm{HP},t}^{\textrm{in}}$ which is the corresponding heat exchanger inlet temperature, either from \ac{hp} to the demand or \ac{hp} to \ac{tess}. Finally, all heatflows are positive and no cooling is considered only space heating. All time-dependent variables are part of the state vector $S_{\text{HP},t}$, except for $\dot{Q}_{\text{HP},t}^{\textrm{TESS}}$ and $\dot{Q}_{\text{HP},t}^{\textrm{D}}$ which are decisions $x_{\text{HP},t}$. The rest of the parameters are in \ref{sec:appA}.}

\subsubsection{Thermal Energy Storage System}

This work considers an underground, perfectly mixed water tank with independent charge and discharge coils as a \ac{tess}. Assuming no mass exchange between the \ac{tess} and the piping system the \ac{tess} thermal balance is:
\begin{equation}
    T_{\text{TESS}, t+1} =  T_{\text{TESS}, t} + \frac{\Delta  t}{m_{\text{TESS}} \cdot c_{\text{TESS}}} \left(  \dot{Q}_{\text{HP},t}^{\textrm{TESS}} - \dot{Q}_{\text{TESS},t}^{\text{D}} - \dot{Q}_{\text{sd}} \right)
    \label{eq:tess_thermal_balance}
\end{equation}

\noindent{where $T_{\text{TESS}, t}$ is the \ac{tess} internal temperature, $\dot{Q}_{\text{sd}}$ is the self-discharge of the \ac{tess}, $m_{\textrm{TESS}}$ is the mass and $c_{\textrm{TESS}}$ is the thermal capacity. Since the optimization horizons $H^{\text{DA/MPC}}$ are 1 or 2 days, the self-discharge $\dot{Q}_{\text{sd}}$ can be assumed to be a constant. To improve on this assumption please refer to \cite{Alpizar-Castillo2024ModellingHouse}.}

Since the tank is assumed to be perfectly mixed its $SoC_{\text{TESS},t}$ is defined by the internal temperature and its limits:
\begin{equation}\label{eq:socTESS}
SoC_{\text{TESS},t}=\frac{T_{\textrm{TESS},t}-\underline{T}_{\textrm{TESS}}}{\overline{T}_{\textrm{TESS}}-\underline{T}_{\textrm{TESS}}}
\end{equation}

Between the \ac{tess}, the building and the \ac{hp} the thermal system has 2 storage devices, 1 heat source and 1 heat sink. The \ac{tess} is an active storage because both its charge and discharge are fully controllable, while the building can store energy in its heat capacity and is discharged exogenously by $\dot{Q}_{\text{loss},t}$ \cite{Risbeck2018Mixed-IntegerSystems}. In combination with the controllable \ac{hp} the heating is a fully flexible carrier. 

\subsection{Electrical carrier}
In the building, all electric devices are connected through power electronic DC/DC converters at a DC bus and connected to the LV grid through a bidirectional AC/DC converter and the building's LV switchboard through an inverter. The power balance in the DC bus is:
\begin{equation}\label{eq:powerBalance}
    P_{\text{PV},t}+P_{\text{BESS},t}+ \gamma_{\text{EV},t}.P_{\text{EV},t}+P_{\text{grid},t}=P_{\text{load},t}+P_{\text{HP},t}\,.
\end{equation}
\noindent{with the left hand side aggregating sources ($P_{\text{PV},t}, P_{\text{grid},t}$) and storage systems ($P_{\text{BESS},t}, \gamma_{\text{EV},t}.P_{\text{EV},t}$) and the right hand side aggregating the controllable ($P_{\text{HP},t}$) and uncontrollable loads ($P_{\text{load},t}$).}

\subsubsection{Battery Energy Storage System}\label{sse:bess}
The remaining devices in the \ac{mces} are all battery-based \ac{ess}. Batteries have complex nonlinear dynamics, and several modelling techniques are presented in the literature \cite{Plett2015, Plett2016, Plett2024BatteryMethods} . In this work, models coming from empirical and physics-based approaches are used, with an \ac{ecm} for performance and a \ac{pbrom} for degradation. Under the \ac{umf}, this is represented in the transition function $\tilde{S}^M_{b,t}(\tilde{S}_{b,t},x_{b,t},\theta_{b,t})$, which contains both the performance model $p_{b,t}^M(\cdot)$ and the ageing model $d_{b}^M(\cdot)$. The performance model predicts stored energy $SoC_{b,t}$ and terminal voltage $v_{t,b, t}$ and parametrized by the parameter vector $\theta_{b,t}$.  The ageing model is used to update the parameters $\theta_{b,t}$ of $p^M_{b,t}(\cdot)$, as in \cite{Slaifstein2025Aging-awareSystems}. The degradation mechanisms to be modelled are \ac{sei} formation and \ac{am} which account for the majority of the battery ageing  \cite{Jin2017, Jin2022, Movahedi2024ExtraMechanism}. This is summarized in:
\begin{subequations}\label{eq:batt}
    \begin{align}
        & S_{b,t+1}=S_{b,t}^M(S_{b,t},x_{b,t},\theta_{b,t})= \begin{cases}
        S_{b,t+1}=p_{b,t}^M \left( S_{b,t},x_{b,t}|\theta_{b,t} \right)\\
         \theta_{b,t+1}=d_{b}^M \left(\theta_{b,t},S_{b,t},x_{b,t}\right)\\
        \end{cases}\\
        & S_{b,t}=\left[SoC,\ i_{R_1},\ OCV_{j},\ z,\ \eta_{k},\ \beta,\ i_{\text{SEI}},\ i_{\text{AM}} \right]^T_{b,t}\\
        & x_{b,t} = P_{b,t},\ y_{b,t} = \left[i, v_t \right]^T_{b,t},\ \theta_{b,t}=Q_{b,t}
    \end{align}
\end{subequations}

The state of the battery $b$ is composed of the state of charge $SoC$, the diffusion current $i_{R_1}$, the open circuit voltages of the electrodes $OCV_j,\ \forall j =n,p$, the Li stochiometry $z$, the \ac{sei} kinetic overpotential $\eta_{k}$, the internal state $\beta$, the \ac{sei} formation rate $i_{\text{SEI}}$, and the active material loss of the electrodes $i_{\text{AM}}$. The decision variable $x_{b,t}$ is $P_{b,t}$, the available measurements $y_t$ are the terminal voltage $v_{t}$ and current $i$, with the ageing parameter $\theta_b$ to be optimized being the cell capacity $Q$. The models $p_{b,t}^M(\cdot)$ and $d_b^M(\cdot)$ are summarized in \ref{sec:appA}. The used model $S_{b,t}^M(\cdot)$ assumes a constant battery temperature $T_{b,t}$ and limited C-rates, allowing us to dismiss Li-plating \cite{Jin2017, Jin2022}.

Eqs. \ref{eq:batt} are combined with the terminal conditions Eqs. \ref{eq:pcDA} and \ref{eq:pcCT} in each corresponding policy. These terminal conditions are at the heart of this paper's contribution; ensuring the turnpike property and leading to the recursive feasibility of the 2-stage non-convex \ac{empc} \cite{Grune2017NonlinearAlgorithms}. Eqs. \ref{eq:pcDA} and \ref{eq:pcCT} bound their terminal sets while ensuring enough flexibility in the controls to not fix the $SoC_{\textrm{BESS}}$ at the beginning of each day. 

\subsubsection{Electric Vehicle}\label{sse:ev}
The final asset in the system is the \ac{ev}, which is modeled as a battery with three particularities: (i) it is not always connected to the system represented in availability $\gamma_{\text{EV},t}$, (ii) while the \ac{ev} is not parked it is being driven consuming $P_{\text{drive,EV}}\sim\mathcal{N}(\mu_d,\sigma_d)$, (iii) at time of departure the user requires it to be at a desired setpoint $SoC_{\text{EV},t_{\text{dep}}} \approx SoC^*_{\text{dep}}$, Eq. \ref{eq:evpen}, as in \cite{Slaifstein2024StochasticHubs}. The complete model can be found in  \ref{sec:appA}.

\subsection{Commentary}
In summary, both policies have four major goals to be fulfilled simultaneously: obtain the best economic outcome $C_{\textrm{grid}}^{\textrm{DA/MPC}}$, with the least degradation $C_{\textrm{loss}}^{\textrm{DA/MPC}}$, while charging the \ac{ev} $p_{\textrm{SoCDep}}^{\textrm{DA/MPC}}$ and maintaining a comfortable inside temperature $p_T^{\textrm{DA/MPC}}$. The first two could be identified as scheduling problems and the second two are soft-tracking problems \cite{Grune2017NonlinearAlgorithms}. The terminal conditions used on the \ac{bess} are used to bound costs $J^{\textrm{DA/MPC}}$, accelerating convergence and avoiding the need for longer horizons \cite{Grune2017NonlinearAlgorithms, Kohler2024AnalysisOverview}. Practically, the terminal constraints are never reached since they always lie outside of the implemented horizon, for both \ac{da} and \ac{mpc}. On the same note, the policy $X_t^{\textrm{MPC}}$ always has to be warm-started with either the \ac{da} prediction (if it's the first of the day) or the previous step prediction $X_{t-1}^{\textrm{MPC}}$. This ensures convergence to local optimality within reasonable times and, more importantly, recursive feasibility \cite{Grune2017NonlinearAlgorithms, KNITRO}.

\section{Case Studies}
\label{sec:results}
\begin{figure}[bt!]
    \centering
    \makebox[\textwidth][c]{\includegraphics[width=1.3\textwidth]{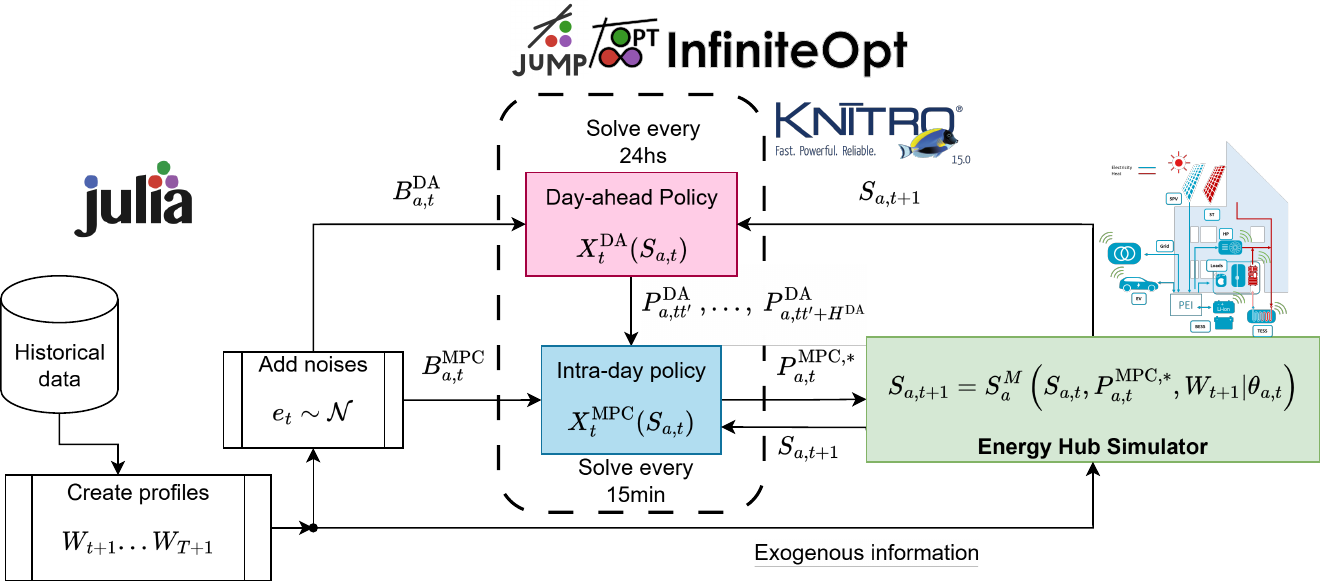}}
    
    \caption{Flow chart of the simulations}
    \label{fig:simflow}
\end{figure}

The system is composed of a 4kWp \ac{spv}, a 15.6kWh \ac{bess}, a 55.6kWh \ac{ev} with \ac{nmc} cells, one 12.5kW \ac{ev} charging point, a 5kWe heat pump, a 200kWh \ac{tess}, a 2.5kWp electrical load, and 17kW LV grid connection. Power consumption profiles ($P_{\textrm{load}}$) were constructed using data from 2021 to 2023 from TU Delft's Green Village smart meter data \cite{GreenVillage}.  The output of the \ac{spv} is taken from \cite{Smets2016, Diab2022, Diab2023}, the market prices $\lambda$ are taken from the EPEX day-ahead and intra-day markets, with $\lambda_{\textrm{sell}}^{\textrm{DA}}=0.95\lambda_{\textrm{buy}}^{\textrm{DA}}$ and $\lambda_{\textrm{sell}}^{\textrm{MPC}}=0.8\lambda_{\textrm{buy}}^{\textrm{MPC}}$ \cite{2023EPEXServices}, and the ambient temperature from \cite{2024WelcomePlatform}. 

The cells used are \ac{nmc} SANYO NCR18650 cells \cite{Jin2022, Planden2022}. The \ac{ecm} was constructed following \cite{Plett2016}. For the thermal models, the parameters are taken from \cite{Damianakis2023Risk-averseConsumption} with the exception of $s_b$ and $r_b$ for the summer, when $s_b=0.1$ and $r_b=0.99$ meaning that the house is properly ventilated and shaded.

The simulations were modelled and run using Julia \cite{julia}, JuMP \cite{JuMP}, and InfiniteOpt \cite{InfiniteOpt}. The chosen solver was KNITRO from Artelys \cite{KNITRO}. All simulations were run using an Intel CPU at 2.60GHz, 4 processors, and 32GB of RAM. Since this is a non-linear non-convex \ac{empc}, global optimality can not be generally guaranteed \cite{Boyd2011ConvexOptimization, Grune2017NonlinearAlgorithms}. Having this in mind, the implementation of the \ac{ems} approaches this limitation by: (i) using warm-starts as initial guess to ensure convergence to the same local optima, (ii) solving day-ahead scheduling with interior point and sequential quadratic programming for the MPC layer, (iii) using the KNITRO tuner and multi-start in each individual step to increase the chance of finding the best local optima \cite{KNITRO}.

The previously mentioned components are combined following the simulation workflow presented in Fig. \ref{fig:simflow}. For each simulation time window $\mathcal{D}_t$, historical data is used to generate different profiles (daily, weekly, etc.) of exogenous information $W_{t+1}$, later passed to the optimization and simulation models. Before being passed to the policies $X^{\text{DA/MPC}}_t(\cdot)$, noise $e_t$ can be added to simulate a forecast. Once the models are built, for each time $t \in \mathcal{D}_t$ the corresponding policies $X^{\text{DA/MPC}}_t(\cdot)$ are solved and evaluated in $S_{a,t}^M(\cdot)$. The simulator feeds back the state $S_{a,t+1}$ to the policies and continues with the loop.

\begin{figure}[b!]
    \centering
    \makebox[\textwidth][c]{\includegraphics[width=1.3\textwidth]{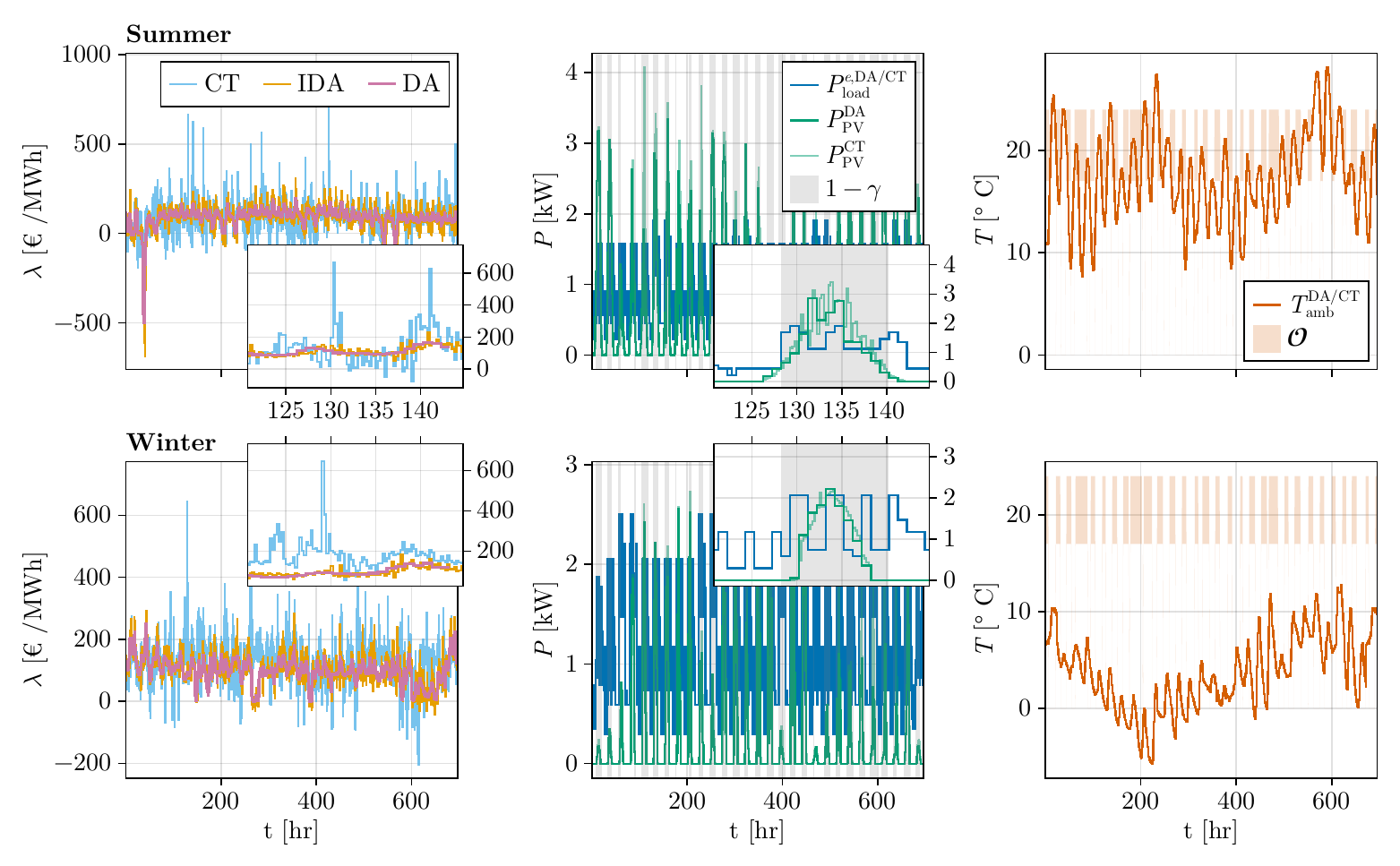}}
    \caption{Inputs for standard months. Summer (top), winter (bottom). From left to right: market prices $\lambda_t$, electric load $P_{\text{load},t}^{\text{e}}$, solar generation $P_{\text{PV},t}$, car availability $\gamma_{\text{EV},t}$, ambient temperature $T_{\text{amb},t}$, and building occupancy $\mathcal{O}_t$.}
    \label{fig:inputsW}
\end{figure}

\subsection{Market participation and operational flexibility}
\label{ssec:cs1_det}

The first contribution of this paper is to challenge common price assumptions in sequential market models. Usually, the literature \cite{Jouini2024PredictiveStudy, Garcia-Torres2021OptimalControl, Li2021IntradaySystem} presents intra-day prices $\lambda^{\textrm{CT}}_t$ as a scaled signal of the day-ahead prices $\lambda^{\textrm{DA}}_t$. However, participating in the intra-day auctions can be beneficial under specific circumstances. These auctions follow a pay-as-clear price $\lambda^{\text{IDA}}_t$ with their own particular dynamics. This work is based on historical prices from the Netherlands, rather than the assumptions in the literature.

Figure \ref{fig:inputsW} presents the standard months for summer and winter of 2023. The first column to the left presents the day-ahead price $\lambda^{\text{DA}}_t$, the intra-day auction price $\lambda^{\text{IDA}}_t$, and the continuous-time intra-day index ID1 $\lambda^{\text{CT}}_t$. This last one is the average price of all transactions in the \ac{ct} intra-day for the past hour. All three prices exhibit distinct frequency harmonics and volatility, contradicting the standard literature assumption \cite{Birkeland2024QuantifyingMarket}. When talking about volatility, this work refers to the standard deviation $\sigma$ and peak values $\lambda_{pk}$ of a given price $\lambda_t$. 

In reality, an operator has to choose how much power is bid in each energy market, with the day-ahead and intra-day auctions being only financial markets. This means that it is not mandatory to dispatch the system following the bids made, but it is mandatory to deliver/receive the contracted euros \texteuro. Not following the promised power dispatch only increases risk exposure in the subsequent balancing markets, which are out of the scope of this paper.

To assess how market participation interacts with the flexibility provided by each asset, different cost function combinations are tested:
\begin{itemize}
    \item $DA \rightarrow DA$: $\pi_{\text{DA}2}$ both cost functions $J^{\textrm{DA/MPC}}$ follow $\lambda^{\textrm{DA}}_t$, with their $\Delta P_g$ evaluated against $\lambda^{\text{CT}}_t$.
    \item $DA \rightarrow CT$: $\pi_{DA \rightarrow CT}$ each cost function $J^{\textrm{DA/MPC}}$ follows $\lambda^{\textrm{DA/CT}}_t$, with $\lambda^{\text{CT}}_t$ being the ID1 index, assuming it's a good approximation of the pay-as-bid mechanism.
    \item $DA \rightarrow IDA$: $\pi_{DA \rightarrow IDA}$ same as $\pi_{DA \rightarrow CT}$ , but the intra-day prices are $\lambda^{\textrm{IDA}}_t$.
    \item $DA+IDA \rightarrow IDA$:  $\pi_{DA2IDA}$ The day-ahead optimization incorporates the dispatch of the \ac{ida}. Thus, the day-ahead dispatch contemplates the \ac{ida} as in $\tilde{J}^{\text{DA}} = \tilde{C}_{\text{grid}}^{\text{DA}}+\tilde{C}_{\text{grid}}^{\text{IDA}} + \tilde{C}_{\text{loss}}^{\text{DA}} + \tilde{p}_{\text{SoCDep}}^{\text{DA}}  + \tilde{p}_{T}^{\text{DA}}$. The $J^{\textrm{MPC}}$ remains the same as in $\pi_{DA \rightarrow IDA}$.
\end{itemize}

Along with the market participation, different sets of flexibility were tested.  Each case is tested under perfect forecast conditions, and thus, the difference in grid costs between each case represents the value of the flexibility provided for the current set of prices. The different cases of flexibility are:
\begin{itemize}
    \item \textit{noFlex}, no flexibility with only PV-HP-EV.
    \item \textit{thFlex}, thermal flexibility with PV-HP-TESS-EV.
    \item \textit{eFlex}, electric flexibility with PV-HP-BESS-EV.
    \item \textit{fullFlex}, multi-carrier flexibility with all PV-HP-BESS-EV-TESS.
\end{itemize}
\noindent{For the cases \textit{noFlex} and \textit{thFlex} the \ac{ev} is on a fast charging mode, i.e. no V2G. This means $\tilde{p}_{\text{SoCDep}}^{\text{DA/MPC}}=w_{SoC}.\gamma_{\text{EV}}.\sum_{t=t'}^{t'+H^{\text{DA/MPC}}}||\tilde{SoC}_{\text{EV},t}^{\text{DA/MPC}}-SoC_{\text{dep}}^*||^2_2.\Delta t$ and the discharging power fixed to $P_{\textrm{EV}}^+=0$.}

\begin{figure}[bt]
    \centering
    \makebox[\textwidth][c]{\includegraphics[width=1.3\textwidth]{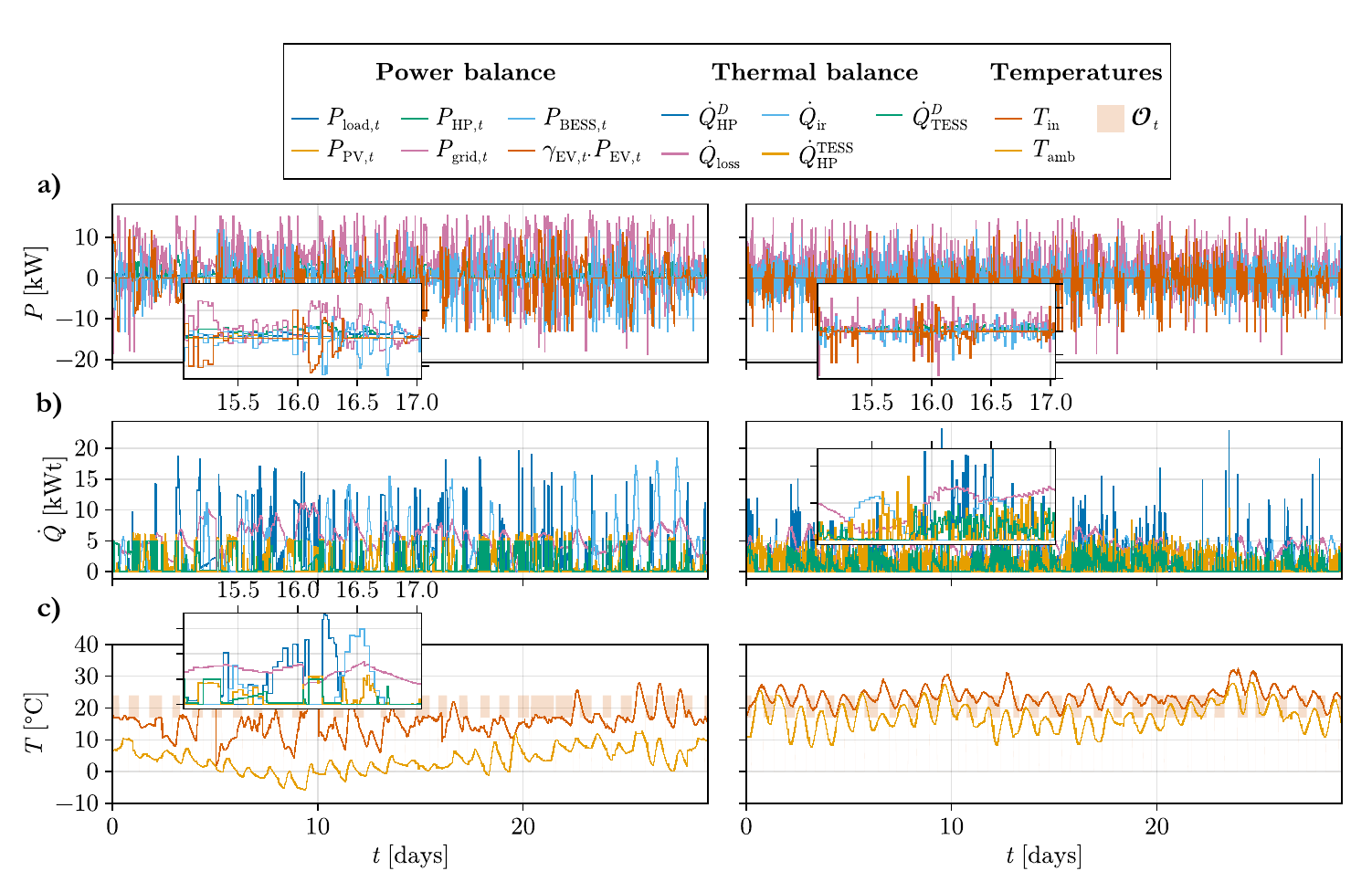}}
    \caption{Summary of the $\pi_{DA2}$ monthly simulations (left) winter and (right) summer. a) Power Balance. b) Heat balance. c) Building temperatures. Inset plots zoom in on days 15 to 17.}
    \label{fig:cs0_ems_fullFlex_da2}
\end{figure}
\begin{figure}[tbh!]
    \centering
    \includegraphics[width=\linewidth]{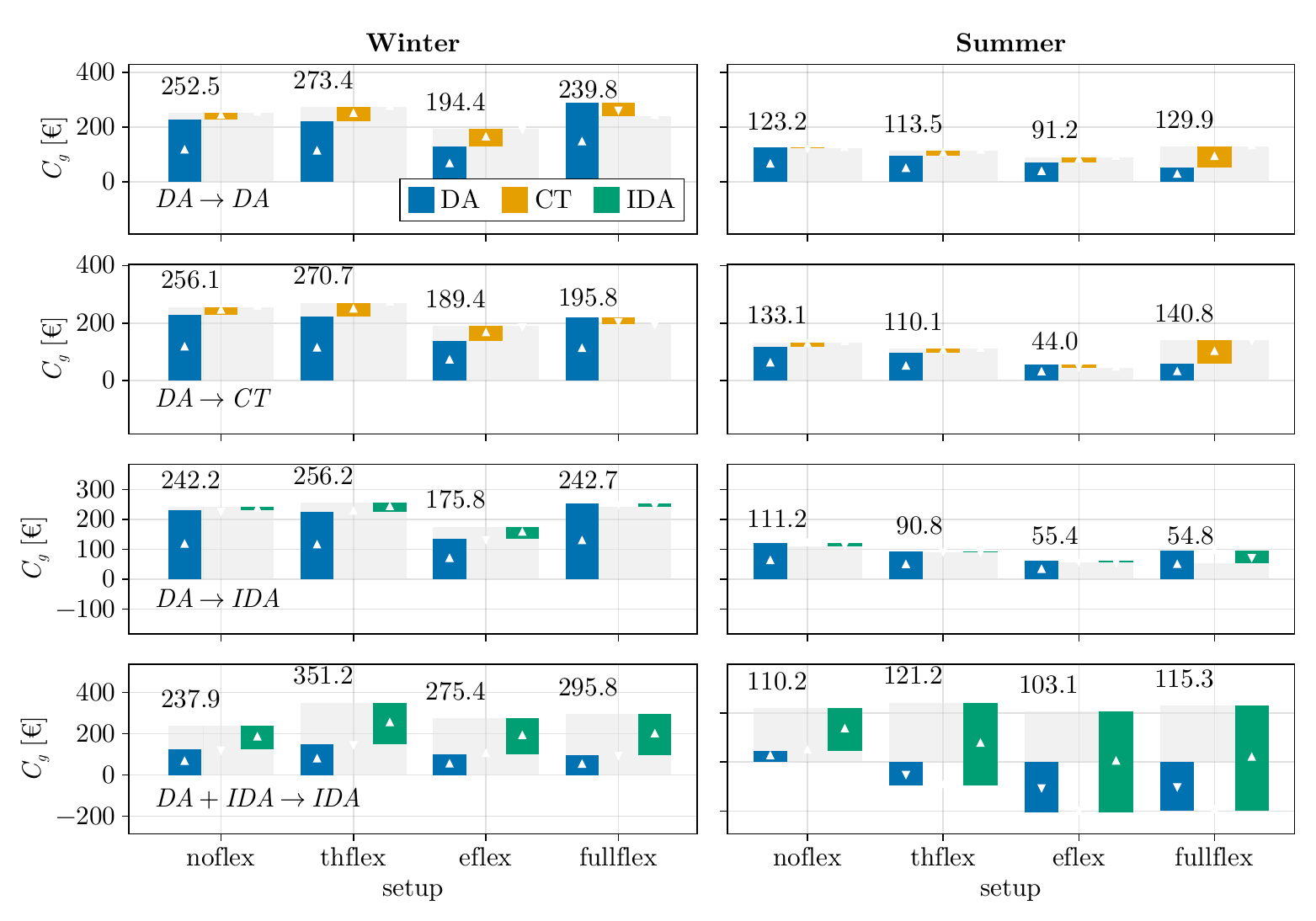}
    \caption{Grid flexibility provided per setup and market. For winter (left), and summer (right).}
    \label{fig:cs1_cgFlex}
\end{figure}

All controllers effectively control the system, maintaining $T_{\text{in},t}$ within bounds, charging the \ac{ev} close to $SoC_{\text{dep}}^*$ at departure, and minimizing its own grid cost $C_{\text{grid}}$. As a representative example, Fig. \ref{fig:cs0_ems_fullFlex_da2} shows the dispatch of the $\pi_{DA2}$ in the \textit{fullFlex} case for the representative months of winter (left) and summer (right). From top to bottom, the first row presents the power balance, the second the heat balance, and the last one the building temperatures. From the power balance, it is clear that the \ac{ev} is the most critical electrical asset due to its capacity and power, followed by the \ac{bess}. The \ac{ev} is correctly charged before departure, and the \ac{bess} is used to arbitrage energy following the frequencies of the $\lambda^{\textrm{IDA}}_t$. Moving down to the heat balance, in summer, the \ac{tess} charges power from solar to avoid heating the house or low-value sells. In winter, the demand is supplied by \ac{hp} and \ac{tess} with the latter oscillating between 50-60\textdegree . On the bottom, the evolution of the $T_{\text{in},t}$ is presented. In summer, the temperature $T_{\text{in},t}$ is close to the upper bound $\overline{T}_{\text{in}}$ when the building is occupied. Whereas, in winter $T_{\text{in},t}$ sits closer to the lower bound $\underline{T}_{\text{in}}$. The house is, in fact, also a passive thermal storage, being heated during low energy prices. This does not always coincide with the building being occupied. The rest of the \textit{fullFlex} results for the remaining policies can be found in \ref{sec:appB}. 

All policies have reasonable computational times $\Delta_{\text{comp}}$ with a mean $\mu_{\Delta}=$7-15s and a standard deviation $\sigma_{\Delta}=$12-47s. The computational time $\Delta_{\text{comp}}$ includes optimization, simulation and utilities, with the simulation having the highest share. More detailed information on the computational load can be found in \ref{sec:appB}.
\begin{figure*}[bh!]
    \centering
    \makebox[\textwidth][c]{\includegraphics[width=1.2\textwidth]{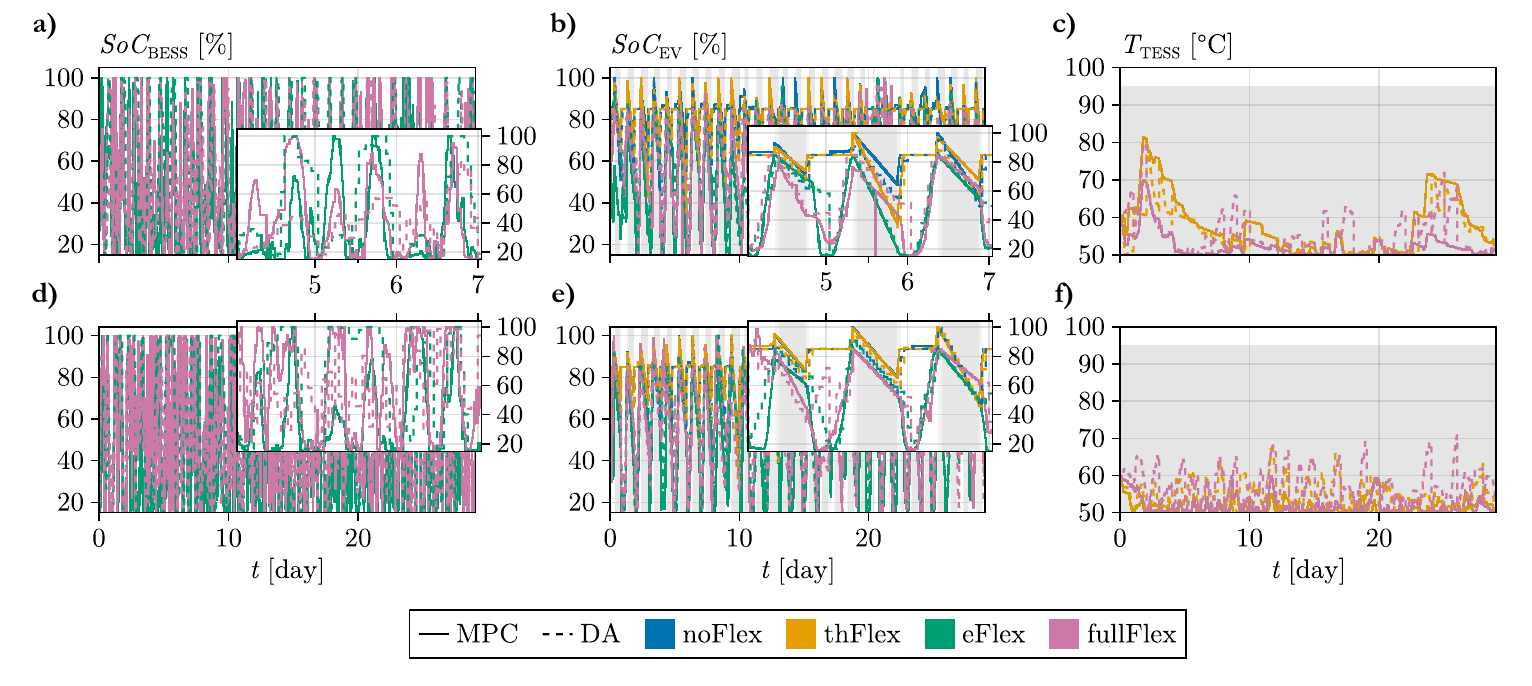}}
    \caption{HESS states under the $\pi_{DA\rightarrow IDA}$ in summer (a-c) winter (d-f) for \ac{da} plan and \ac{mpc} realization. a,d) BESS State of Charge, b,e) EV State of Charge and c,f) TESS Temperature}
    \label{fig:cs1_hess_daida}
\end{figure*}

Figure \ref{fig:cs1_cgFlex} presents the grid costs $C_{\text{grid}}$ of each flexibility setup for each policy $\pi$. When comparing across markets and flexibilities, there is no policy $\pi$ that comes up on top across flexibilities and seasons. The predictive policy $\pi_{DA2IDA}$ achieves the lowest $C_{\text{grid}}$ for the \textit{noflex} setup. The \textit{thflex} setup provides flexibility only during summer for the first three policies, while during the winter it adds $C_{\text{grid}}$. The \textit{eFlex} setup achieves the lowest grid cost for all policies $\pi$ with the exception of $\pi_{DA \rightarrow IDA}$ in the summer.

Particularly, the $\pi_{DA2IDA}$ policy has the best day-ahead costs $C_{\text{grid}}^{\textrm{DA}}$ due to offering $P_{\textrm{grid},t}^{\textrm{DA}}$ beyond what is physically available to later take losses in the $C_{\text{grid}}^{\text{IDA}}$. Since in $\pi_{DA2IDA}$ the grid power is $P_{\textrm{grid},t}=P_{\textrm{grid},t}^{\textrm{DA}}+P_{\textrm{grid},t}^{\textrm{IDA}}$, with its corresponding costs, and $\lambda_t^{\text{DA}}>\lambda_t^{\text{IDA}}$ most of the time the policy chooses to maximize $P_{\text{grid},t}^{\text{DA},-}$  and maintain a reasonable overall $P_{\text{grid},t}$. Later, in the $X^{\text{MPC}}_t$ grid power is $P_{\textrm{grid},t}=P_{\textrm{grid},t}^{\textrm{IDA}}$ warm-started with the sequence decided by $X^{\text{DA}}_{t-1}$, leading to high intra-day auction cost $C_{\text{grid}}^{\text{IDA}}$ for the system (monetary losses). In other words, the policy $\pi_{DA2IDA}$ tries to game the two markets, \ac{da} and \ac{ida}, by predicting the price auctions $\lambda_t^{\text{IDA}}$ in the first stage but fails when the dataset is very diverse (high price volatility and large ambient temperature variations). In preliminary weekly simulations where price volatility and temperature variations were limited, this policy was the top performer.

Continuing, not all policies $\pi$ can decrease $C_{\text{grid}}$ when passing from \textit{eFlex} to \textit{fullFlex}, and even when they do, the change is marginal. The \textit{fullFlex} setup only achieves lower $C_{\text{grid}}$ in the summer $\pi_{DA \rightarrow IDA}$ because it already saves costs in the \textit{thflex} setup. The \ac{tess} by itself only provides short-term flexibility in the first three policies of the summer. The main limitations are a combination of: (i) the \ac{tess} maximum heat-flow $\overline{\dot{Q}}_{\text{TESS}}=5kWth$ is smaller than the \ac{hp} $\overline{\dot{Q}}_{\text{HP}}=20kWth$, (ii) the complementarity between heating the \ac{tess} or the building, Eq. \ref{eq:hp}d, and (iii) the $H^{\text{MPC}}$ not being long enough to capture the value of the \ac{tess}. These will be further explained in the following sections.

The rationale behind the effectiveness of \textit{eFlex} is that the volatility of $\lambda_{t}^{\text{DA/CT}}$ allows the \ac{bess} to arbitrage energy throughout the day, capturing fast price spikes and dips. The total grid cost significantly decreases when the \ac{bess} is introduced and the bidirectional charging is allowed (\textit{eFlex} and \textit{fullFlex}). Summarizing, the worst performing policy seems to be $\pi_{DA2}$, unaware of the \ac{ida} and \ac{ct} markets. Integrating the \ac{tess} and the electric storage unlocks value only under specific conditions. The only policy that ensures the synergy between \ac{tess} and \ac{ess} is the $\pi_{DA\rightarrow IDA}$. The statistical  validation of the \ac{tess} short-term flexibility is further explored in Section \ref{ssec:cs2_noisy}.

\subsection{HESS operation: from plan to execution}\label{ssec:hess}

Moving to the \ac{hess} states, there are relevant differences between the \ac{da} plans $X^{\textrm{DA}}_t$ and the implemented \ac{mpc} actions $X^{\textrm{MPC}}_t$. The $\pi_{DA \rightarrow IDA}$ is used as a representative example in Fig. \ref{fig:cs1_hess_daida}. From left to right, the $SoC_{\textrm{BESS},t}$ , Fig \ref{fig:cs1_hess_daida}a,d, changes its periodicity from \ac{da} to \ac{mpc}, due to the higher frequency component of the intra-day prices $\lambda^{\textrm{IDA/CT}}_t$. This is measured in full equivalent cycles $FEC_{\textrm{BESS}}$ which increase between 30-60\% in the winter depending on the flexibility setup. The inset presents a close-up of the end of the first week, to appreciate the difference between $\tilde{SoC}_{\textrm{BESS},t}^{\textrm{DA}}$ and $SoC_{\textrm{BESS},t}$. The peaks and cycles planned by $X^{\text{DA}}_t$ are replaced by irregular charge/dishcarge cycles to compensate for the variations in $W_{t+1}$ (mainly $\lambda_{t}^{\text{MPC}}$ and $P_{\text{PV},t}$) and the incoming information at the tail of the horizon $H^{\textrm{MPC}}+1$. The change in frequency from \ac{da} to \ac{mpc} affects the other storages with less notorious effects. The impact on \ac{ev} and \ac{tess} is limited because both are tied to other demands (mobility and heating). Even though the \ac{ev} is the preferred \ac{ess} due to its high round-trip efficiency and large capacity, it can not compensate for the short-term variations because it's not always connected and it has to be charged before departure. Thus, for the \ac{bess} there are few qualitative differences between \textit{eFlex} and \textit{fullFlex}, with the biggest impact coming from the change between \ac{da} to \ac{mpc}. For the \ac{ev} the trajectories are consistent throughout seasons, and policies.
\begin{figure}[bt!]
    \centering
    \makebox[\textwidth][c]{\includegraphics[width=1.3\textwidth]{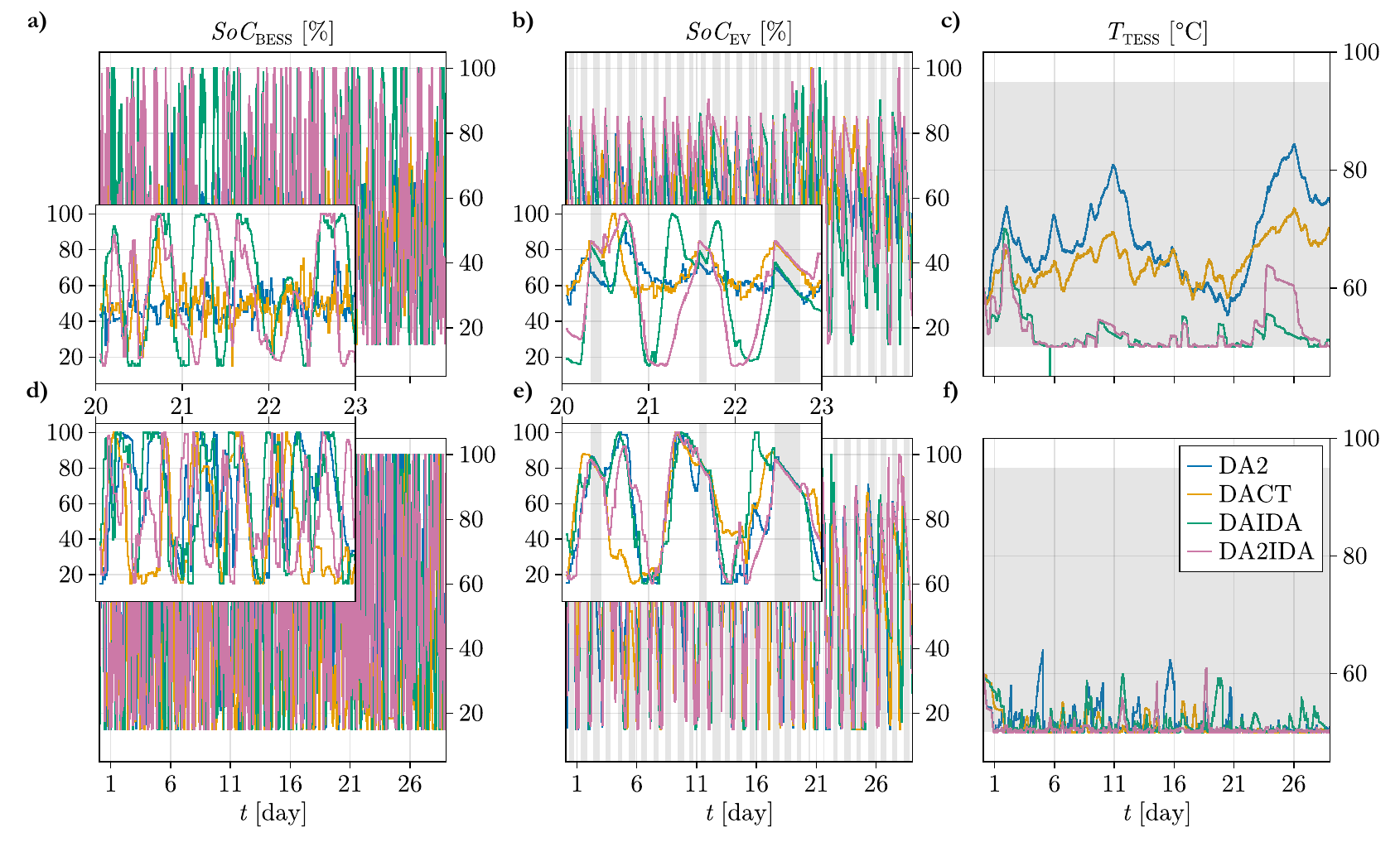}}
    \caption{Monthly HESS dispatch in the \textit{fullFlex} case. Summer (a-c), winter (d-f), BESS State of Charge (a,d), EV State of Charge (b,e), TESS Temperature (c,f).}
    \label{fig:cs1_HESS_fullFlex}
\end{figure}

Finally, the \ac{tess} reduces its flexibility supply from \ac{da} to \ac{mpc}, quantified by the decrease in full equivalent cycles $FEC_{\textrm{TESS}}$. This is partially due to the shorter horizon $H^{\textrm{MPC}}$ w.r.t. $H^{\textrm{DA}}$, resulting in a greedier policy for the \ac{tess}. Greedier, meaning discharging without recharging. In the summer, the $\tilde{T}^{\textrm{DA}}_{\textrm{TESS},t}$ captures the negative prices from $\lambda^{\textrm{DA}}_t$, which is mitigated in $T_{\textrm{TESS},t}$. As mentioned earlier, although temperature-based models offer a computational advantage over flow-based models, they do not completely address the early depletion of thermal storage. The reasons for this are three: (i) the shorter the opt. horizon $H$ the less attractive is the flexibility of the \ac{tess} (low $\eta$, high $Q$, high $\dot{Q}_{\text{sd}}$) \cite{Prat2024HowProblems, Darivianakis2017AManagement}, (ii) the thermal setup (complementarity and series connection) and (iii) the fact that \ac{tess} flexibility mostly causes $C_{\textrm{grid}}$ savings since there is no heat-to-power and negative prices can be captured by other \ac{ess} or the building. This is a shortcoming of the chosen policy ($H^{\text{MPC}}$) and the design assumptions (series connection and complementarity) since some form of seasonal information could inform daily decisions or a change in the setup design, and it is an open point for further research \cite{Darivianakis2017AManagement}. Summarizing, from \ac{da} to \ac{mpc} the increase in charge-discharge cycles is absorbed by the electrical storage (\ac{bess} and \ac{ev}) and decreased in the \ac{tess}.

On a broader scope for all market sequences, the \ac{hess} operation in the \textit{fullFlex} case is presented in Fig. \ref{fig:cs1_HESS_fullFlex}. All the different price signals $\lambda_t$ have different frequency spectra, with the \ac{ct} having the largest volatility, followed by \ac{ida} and \ac{da} last. As the \ac{ev} and \ac{tess} are constrained, the high efficiency \ac{bess} is always available to accommodate changes in $W_{t+1}$.
For the \ac{ev}, the main difference is the moments in which power is discharged. All policies have roughly the same $FEC_{\text{EV}}$, with the difference being less than a cycle. However, when zooming into the end of the 3rd week, in summer, policies $\pi_{DA2}$ and $\pi_{DA\rightarrow CT}$ produce smaller $SoC_{b,t}$ deviations than the $\pi_{DA \rightarrow IDA}$ and $\pi_{DA2IDA}$ policies, Figs. \ref{fig:cs1_HESS_fullFlex}a,b. On the thermal side, the \ac{tess} temperature in winter is brought down to its lower limit and flexibly operates within the bottom 10\textdegree C, Figure \ref{fig:cs1_HESS_fullFlex}f, similar to the results presented in \cite{Slaifstein2025Aging-awareSystems}. However, in Figure \ref{fig:cs1_HESS_fullFlex}c $T_{\text{TESS},t}$ trajectories of policies  $\pi_{DA2}$ and $\pi_{DA\rightarrow CT}$ are consistently higher than the sequential policies trajectories, causing their high $C_{\text{grid}}$ in Figure \ref{fig:cs1_cgFlex}.

The change in total $FEC$ of the \ac{ess} from \ac{da} to \ac{mpc} for all flexibilities and market combinations is presented in Table \ref{tab:cs1_FEC}. In the \textit{noFlex} and \textit{thFlex} cases, the $\Delta FEC=\Delta FEC_{\text{EV}} \leq 10\%$ meaning there is no relevant difference in the \ac{ev} cycles from \ac{da} to \ac{mpc}. Significant variation is introduced in \textit{eFlex} and \textit{fullFlex} with the \ac{bess} and the \ac{ev} smart bidirectional charging. The difference between $\lambda_t^{\text{DA}}\rightarrow \lambda_t^{\text{CT}}$ and $\lambda_t^{\text{DA}}\rightarrow \lambda_t^{\text{IDA}}$ increases in the summer, leading to an increase of more than 50\% in all \textit{eFlex} cases. Whereas in winter, the change in Ah processed is less than 10\%. In the \textit{fullFlex} setups the throughput increases between 16\% to almost 50\%. This is because the \ac{tess} does not realize the $FEC$ expected by $X^{\text{DA}}_t$, meaning the \ac{bess} and \ac{ev} have to compensate for it. In general, the increase from \ac{da} to \ac{mpc} realization is due to the higher short-term volatility of $\lambda_t^{\text{CT}}$ and $\lambda_t^{\text{IDA}}$ with respect to $\lambda_t^{\text{DA}}$. The worst case is $\pi_{DA\rightarrow CT}$ in \textit{eFlex} during the summer with almost a 67\% increase, due to the increased volatility from $\lambda_t^{\text{DA}}\rightarrow \lambda_t^{\text{CT}}$. But how does this increased throughput impact battery degradation? A deeper analysis of this is presented in the following Section \ref{ssec:battDeg}.
\begin{table}[b!]
\centering
\begin{adjustbox}{width=\textwidth}
\begin{tabular}{lllllllll}
\hline
                  & \multicolumn{4}{c}{\textbf{Winter}}                                                                                                          & \multicolumn{4}{c}{\textbf{Summer}}                                                                                                    \\
                  & \textbf{DA2}                          & \textbf{DACT}                & \textbf{DAIDA}               & \textbf{DA2IDA}               & \textbf{DA2}                  & \textbf{DACT}                 & \textbf{DAIDA}                & \textbf{DA2IDA}               \\
\hline
\textbf{noflex}   & \cellcolor[HTML]{B0C7E4}-6\% & \cellcolor[HTML]{AEC5E3}-6\% & \cellcolor[HTML]{A8C0E1}-6\% & \cellcolor[HTML]{7DA2D2}-10\% & \cellcolor[HTML]{6391C9}-12\% & \cellcolor[HTML]{759DCF}-10\% & \cellcolor[HTML]{6C97CC}-11\% & \cellcolor[HTML]{5A8AC6}-13\% \\
\textbf{thflex}   & \cellcolor[HTML]{DFE8F5}-2\% & \cellcolor[HTML]{D9E3F2}-2\% & \cellcolor[HTML]{C7D6EC}-4\% & \cellcolor[HTML]{CCDAEE}-3\%  & \cellcolor[HTML]{C4D5EB}-4\%  & \cellcolor[HTML]{B9CDE7}-5\%  & \cellcolor[HTML]{BDCFE8}-5\%  & \cellcolor[HTML]{B8CCE7}-5\%  \\
\textbf{eflex}    & \cellcolor[HTML]{FCF7FA}3\%  & \cellcolor[HTML]{FCF4F7}4\%  & \cellcolor[HTML]{FCE8EB}10\% & \cellcolor[HTML]{FCF0F3}6\%   & \cellcolor[HTML]{FA8F91}50\%  & \cellcolor[HTML]{F8696B}67\%  & \cellcolor[HTML]{F98587}54\%  & \cellcolor[HTML]{FA9597}47\%  \\
\textbf{fullflex} & \cellcolor[HTML]{FBC1C3}27\% & \cellcolor[HTML]{FA989A}46\% & \cellcolor[HTML]{FA9294}48\% & \cellcolor[HTML]{FBCDD0}22\%  & \cellcolor[HTML]{FCD9DC}16\%  & \cellcolor[HTML]{FBBBBD}30\%  & \cellcolor[HTML]{FA989A}46\%  & \cellcolor[HTML]{FA9698}47\% \\
\hline
\end{tabular}
\end{adjustbox}
\caption{Change in $FEC$ from DA plan to MPC realization for all markets and flexibilities.}
\label{tab:cs1_FEC}
\end{table}

\subsection{Battery degradation}\label{ssec:battDeg}

\begin{figure}[bt]
    \centering
    \makebox[\textwidth][c]{\includegraphics[width=1.3\textwidth]{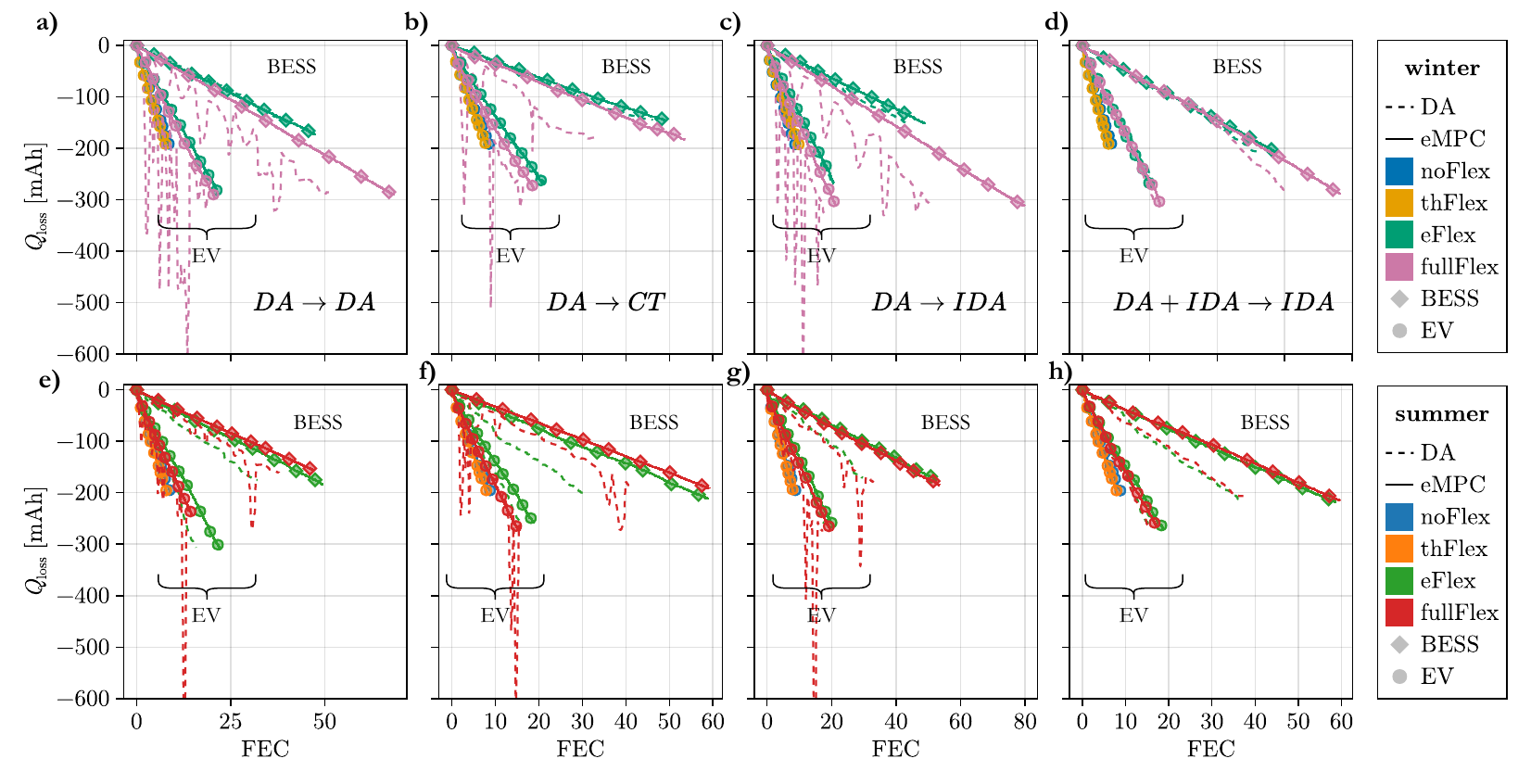}}
    \caption{Degradation trajectories for the different policies (from left to right), winter (a-d), summer (e-h).}
    \label{fig:cs1_deg}
\end{figure}

Regarding battery degradation,  Fig. \ref{fig:cs1_deg} summarizes the impact of the different policies on the \ac{ess}. It presents the lost capacity $Q_{\text{loss},b,t}$ as a function of $FEC_{b,t}$ where $b \in [\text{BESS, EV}]$. An effective policy $\pi$ would steer the trajectories to the top-right, maximizing $FEC_{b,t}$ with minimum $Q_{\text{loss},b,t}$. On the contrary, an ineffective policy would steer the system towards the bottom-left corner, maximum degradation with minimum throughput. In general, the \ac{da} predictions are always below the \ac{mpc} realization $\tilde{Q}_{\textrm{loss},T}^{\textrm{DA}} < Q_{\textrm{loss},T}$ due to the increase in $\tilde{FEC}_{b,T}^{\textrm{DA}} < FEC_{b,T}$. The transition from \textit{eFlex} to \textit{fullFlex} leads to different results depending on the season. In the winter, all \textit{fullFlex} \ac{bess} trajectories have steeper degradation control ($\frac{\partial Q_{\text{loss,BESS},t}}{\partial FEC_{\text{BESS},t}}$), whereas in the summer this only happens with $\pi_{DA \rightarrow IDA}$. For the other three policies in the summer, the slope decreases, meaning that degradation control improves with the addition of the \ac{tess}. The $Q_{\text{loss,EV},t}$ value is roughly the same for \textit{eFlex}  and \textit{fullFlex} across different seasons and policies. Coincidentally, in the summer Figures \ref{fig:cs1_deg}e-h,  the addition of the \ac{tess} leads to higher $C_{\text{grid}}$ but also to lower $Q_{\textrm{loss}}$ in all policies \textbf{except for $\pi_{DA\rightarrow IDA}$ which lowers its grid cost but loses more capacity}. This is because the addition of the \ac{tess} sometimes causes the solver to find a different local minima where the \ac{bess} is cycled less, reducing its cyclic ageing. The downside of this is that in those occasions the flexibility is reduced (i.e., an increase in $C_{\text{grid}}$ when \textit{eFlex} $\rightarrow$ \textit{fullFlex}). Quantitatively, the policies with the least degradation per cycle (smallest $\frac{\partial Q_{\text{loss},t}}{\partial FEC_{t}}$) are the $\pi_{DA2}$ and $\pi_{DA2IDA}$ ranging from \textbf{6.4 to 6.9 mAh per eq. cycle}, which coincidentally had the highest grid costs $C_{\text{grid}}$.

Overall, the cause of the degradation control is the ageing \ac{pbrom}, because it directly relates battery current $i_{b,t}$ with the capacity fade $Q_{\text{loss},b,t}$ in a dynamic manner. This creates schedules where the cycles capture value while protecting battery lifetime \cite{VegaGarita2025TheReview, Plett2024BatteryMethods, Li2023EnsembleManagement}. Another relevant phenomenon is the fact that since the batteries are at the beginning of their lifetime, the calendar ageing factor of the SEI model is dominant versus the cyclic component. This leads the policies $\pi$ to increase $FEC$ since the final $Q_{\text{loss}}$ will stay roughly the same. In other words, the \ac{bess} and \ac{ev} will degrade anyway, so it's better to cycle them. Flexibility and degradation control meet at Table \ref{tab:cs1_DeltaCgQloss}, which presents the ratio between the flexibility gained $\Delta C_{\text{grid}}$ and the total capacity lost $Q_{\text{loss}}$, i.e., how much flexibility was lost/gained by each lost mAh of storage capacity. Even though the \textit{eFlex} cases of the sequential policies $\pi_{DA \rightarrow CT}$ and $\pi_{DA \rightarrow IDA}$ spend their mAh more wisely than their counterparts, the policies with the best degradation control are $\pi_{DA2}$ and $\pi_{DA2IDA}$.

\begin{table}[bt]
\centering
\begin{adjustbox}{width=\textwidth}
\begin{tabular}{lrrrrrrrr}\hline
                  & \multicolumn{4}{c}{\textbf{Winter}}                                                                                                                      & \multicolumn{4}{c}{\textbf{Summer}}                                                                                                                      \\
                  & \multicolumn{1}{l}{\textbf{DA2}} & \multicolumn{1}{l}{\textbf{DACT}} & \multicolumn{1}{l}{\textbf{DAIDA}} & \multicolumn{1}{l}{\textbf{DA2IDA}} & \multicolumn{1}{l}{\textbf{DA2}} & \multicolumn{1}{l}{\textbf{DACT}} & \multicolumn{1}{l}{\textbf{DAIDA}} & \multicolumn{1}{l}{\textbf{DA2IDA}} \\
                  \hline
\textbf{noflex}   & \cellcolor[HTML]{FCFCFF}-        & \cellcolor[HTML]{FCFCFF}-         & \cellcolor[HTML]{FCFCFF}-          & \cellcolor[HTML]{FCFCFF}-           & \cellcolor[HTML]{FCFCFF}-        & \cellcolor[HTML]{FCFCFF}-         & \cellcolor[HTML]{FCFCFF}-          & \cellcolor[HTML]{FCFCFF}-           \\
\textbf{thflex}   & \cellcolor[HTML]{F8696B}-0,11    & \cellcolor[HTML]{F99396}-0,07     & \cellcolor[HTML]{FCFCFF}0,00       & \cellcolor[HTML]{FBFCFE}0,00        & \cellcolor[HTML]{D6EDDE}0,05     & \cellcolor[HTML]{A2D8B1}0,11      & \cellcolor[HTML]{ABDCB9}0,10       & \cellcolor[HTML]{F9B1B3}-0,05       \\
\textbf{eflex}    & \cellcolor[HTML]{97D3A8}0,13     & \cellcolor[HTML]{7CC891}0,16      & \cellcolor[HTML]{7DC992}0,16       & \cellcolor[HTML]{F98F91}-0,08       & \cellcolor[HTML]{C8E7D3}0,06     & \cellcolor[HTML]{63BE7B}0,19      & \cellcolor[HTML]{97D3A8}0,13       & \cellcolor[HTML]{F1F8F5}0,01        \\
\textbf{fullflex} & \cellcolor[HTML]{EBF5F0}0,02     & \cellcolor[HTML]{94D2A5}0,13      & \cellcolor[HTML]{FBFBFE}-0,00      & \cellcolor[HTML]{F87577}-0,10       & \cellcolor[HTML]{FBE4E7}-0,02    & \cellcolor[HTML]{FBE5E8}-0,02     & \cellcolor[HTML]{99D4AA}0,12       & \cellcolor[HTML]{FBEDF0}-0,01      
\\ 
\hline
\end{tabular}
\end{adjustbox}
\caption{Grid cost savings per unit of lost battery capacity [\texteuro/mAh] for all markets and flexibilities relative to their \textit{noflex} setup.}
\label{tab:cs1_DeltaCgQloss}
\end{table}
\begin{figure}[bt]
    \centering
    \includegraphics[width=0.8\linewidth]{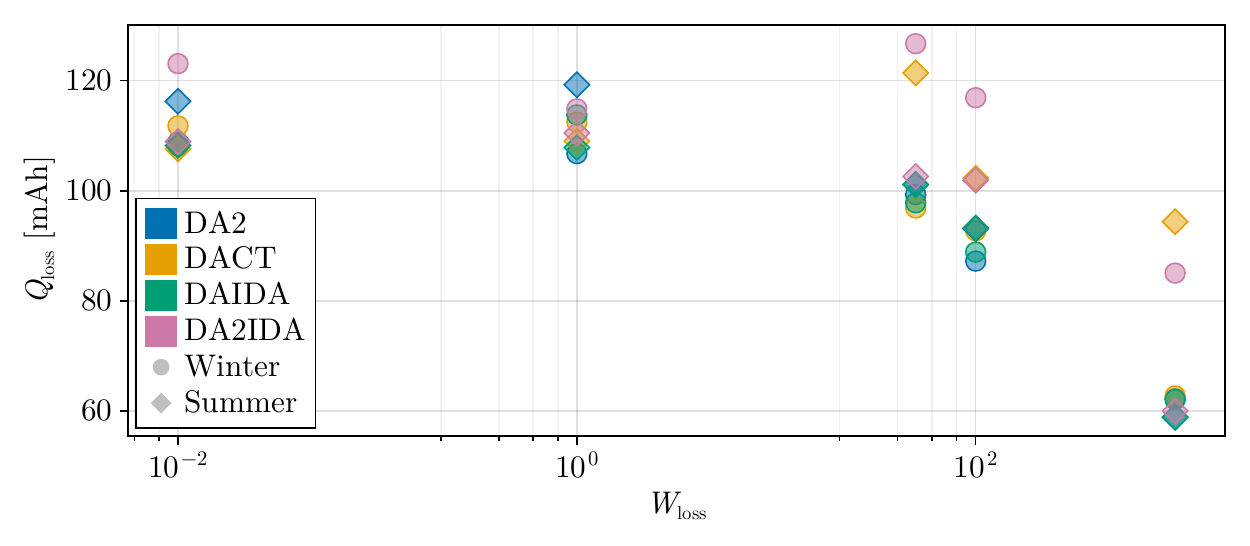}
    \caption{Total capacity fade $Q_{\text{loss}}$ of \textit{eFlex} under different policies $\pi$ for standard weeks ($T=1$ week).}
    \label{fig:DegControl}
\end{figure}

The total degradation is controllable by means of $w_{\text{loss}}$. As an example, Figure \ref{fig:DegControl} presents the sensitivity of the total capacity fade $Q_{\text{loss}}$ against $w_{\text{loss}}$ for the \textit{eFlex} setup in standard weeks ($T=7$ days) of summer and winter. As expected, as $w_{\text{loss}}$ increases the capacity fade is reduced. The pattern is nonlinear, which is natural given that the batteries used are at the beginning of their lifetime and that the optimization problem is numerically scaled at each timestep. This leads to a range ($w_{\text{loss}} \in [0.01, 10]$) where $Q_{\text{loss},T}$ is roughly constant, where the calendar ageing dominates with respect to the cyclic ageing, since the former depends on $\sqrt{t}$. This calendar component presents a higher slope w.r.t. the cyclic components of $i_{\text{SEI},t}$ and $i_{\text{AM},t}$. As the weight increases past the breaking point, the capacity fade is reduced by decreasing the $SoC_{b,t}$ as much as possible and by reducing the \ac{ev} V2G service. As a caveat, since the knee-point is around $w_{\text{loss}}=500$ the impact on the other objectives, it is not straightforward. As $w_{\text{loss}}$ increases, it becomes comparable to the thermal comfort penalty $p_{T}$ and the \ac{ev} charging penalty $p_{\text{SoCDep}}$. Thus, a priori, the increase in $w_{\text{loss}}$ only ensures a decrease in $Q_{\text{loss}}$, not a direct increase in grid cost $C_{\text{grid}}$. For example, for $w_{\text{loss}}=1000$ there are local solutions with extended battery lifetime and relatively low grid cost, but at the expense of not charging the \ac{ev} or heating the house. Thus, for each new $w_{\text{loss}}$ a new set of weights for the penalties needs to be found, as it is usual in optimization problems with weighted-sum objectives. As the calendar degradation component naturally decreases, the sensitivity range becomes wider and reduces this effect. For further details, please refer to \cite{Movahedi2024ExtraMechanism, Jin2022, Dorronsoro2025BatteryModels}.

\subsection{Noisy performance}\label{ssec:cs2_noisy}
Finally, the same \textit{fullFlex} case is tested under noisy conditions at a noise level of 10\% for the summer and winter standard weeks. Each policy is tested under $N_s=50$ realizations of the exogenous processes $W_{t+1} \in \Omega$. Each forecast realization/trajectory is defined as $\tilde{B}_{tt'+1,\omega}=W_{t+1}+\varepsilon_{t,\omega}$ with $\omega \in [1,N_s]$ being the realization index and $\varepsilon_{t,\omega} \sim \mathcal{N}(0,\sigma)$ a white noise with standard dev./noise level $\sigma$. In summary, at each timestep $t$ the policy $\pi$ receives a noisy forecast $\tilde{B}_{tt'+1,\omega}$ optimizes over $t'$ and is evaluated in the simulator under the test data $W_{t+1}$.

\begin{figure}[bt!]
    \centering
    \includegraphics[width=\linewidth]{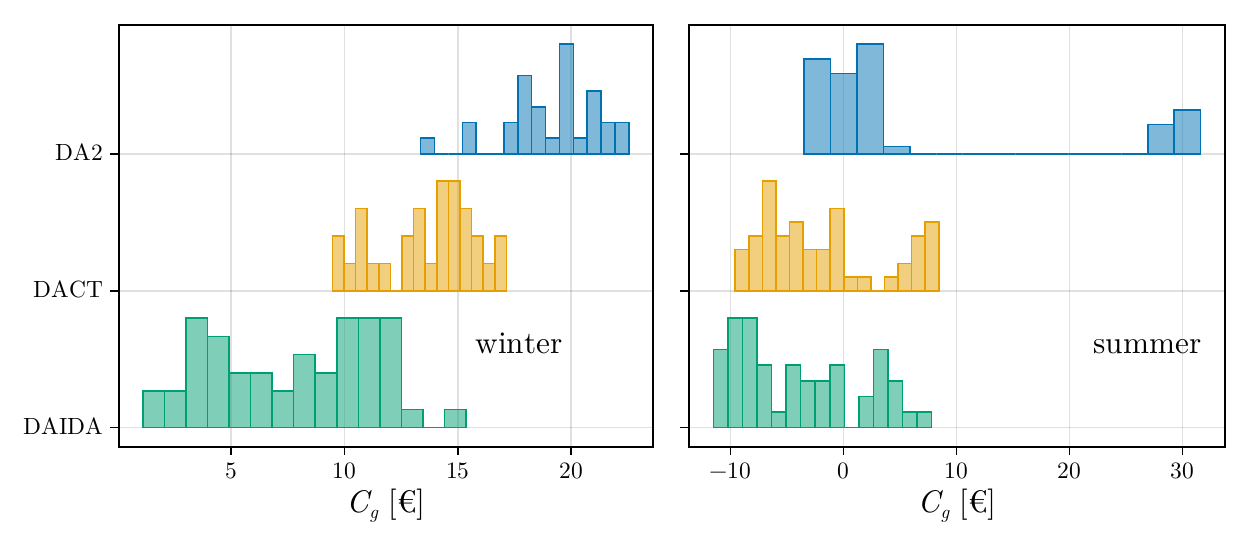}
    \caption{Grid costs distributions of \textit{fullfFlex} under noisy conditions for weekly simulations.}
    \label{fig:distCg}
\end{figure}

The grid cost distributions are shown in Fig. \ref{fig:distCg} for the \textit{fullFlex} case under all policies, except for the $\pi_{DA2IDA}$ since it has proven to become fragile under diverse operating conditions. From Figure \ref{fig:distCg}, we see that $\pi_{DA2}$ is the riskiest policy, having the most cases with higher $C_{\text{grid}}$. The best policy seems to be $\pi_{DA \rightarrow IDA}$ in both seasons, with the distribution further to the left. This is due to the particular statistics of each market signal. Since the original $\lambda^{\text{DA}}_t$ degrades the most when introducing noise $\varepsilon_{t,\omega}$, following the noisy signal $W_{t+1}+\varepsilon_{t,\omega}$ results in the worst performance. On the other end, the $\lambda^{\text{IDA}}_t$ is usually below the $\lambda_t^{\text{DA}}$ , thus the cases where the noisy forecast overestimates \ac{da} and underestimates \ac{ida} result in unexpected gains for the $\pi_{DA\rightarrow IDA}$ because the spread between IDA and DA is smaller than expected, allowing the policy to buy lower than expected in the DA market and sell higher than expected in the IDA.

\begin{figure}[tb]
    \centering
    \includegraphics[width=\linewidth]{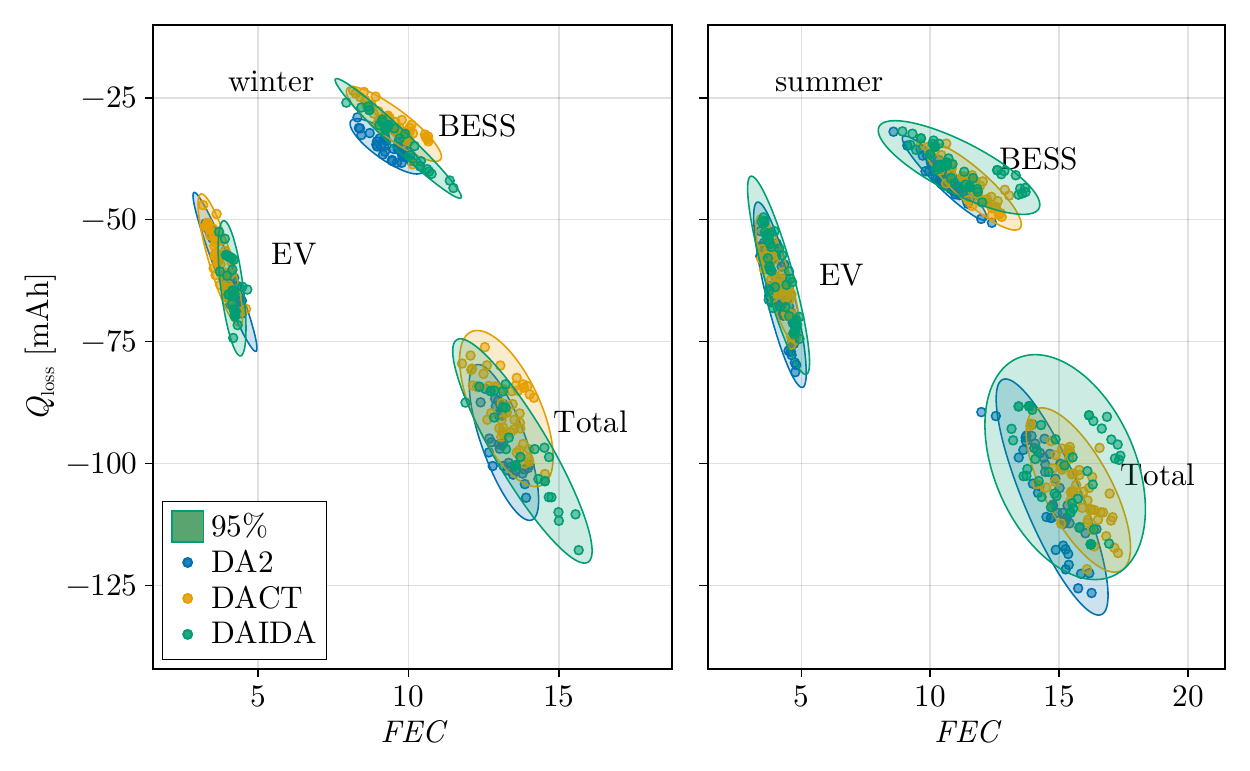}
    \caption{Final degradation distributions for each policy under noisy conditions.}
    \label{fig:distDeg}
\end{figure}

The impact of uncertainty on the degradation of the battery storage depends on the market sequence, presented in Figure \ref{fig:distDeg}. For the \ac{ev}s the $FEC$ are tightly coupled to the true $P_{\text{drive},t}$ and $\gamma_{\text{EV},t}$, as such the small variation on $FEC_{\text{EV},T}$ explained on the noisy forecast $\hat{P}_{\text{drive},t}$ and $\hat{\gamma}_{\text{EV},t}$ that generates volatility on the $SoC_{\text{EV},t}$. The closer the $SoC_{\text{EV},t}$ is to its lower bound $\underline{SoC}_{\text{EV}}$ the smaller the capacity fade $Q_{\text{loss,EV},T}$. In other words, incorrectly forecasting mobility demand leads to low $SoC_{\text{EV},t}$ results in smaller degradation, but at the risk of running out of battery in the middle of a trip. On the other hand, the \ac{bess} has almost a $\Delta FEC_{\text{BESS},T}\approx50\%$ variation due to poor forecasting. However, as opposed to the \ac{ev}, since the \ac{bess} is always available, the policies achieve a lower $\frac{\partial Q_{\text{loss}}}{\partial FEC}$/improved degradation control. When comparing the policies to each other, the policy $\pi_{DA2}$ has the overall worst degradation control/steepest $\frac{\partial Q_{\text{loss}}}{\partial FEC}$ in both seasons. Finally, there is no significant difference between the other two policies, with the $\pi_{DA\rightarrow IDA}$ being better positioned since its distribution of $C_{\text{grid}}$ has the lowest value at risk.

Summarizing, in this Section, it has been shown how the proposed novel two-layer \ac{empc} participates sequentially in the day-ahead and intra-day markets. From the power balance, the \ac{ev} and the \ac{bess} are the most critical electrical assets. While the \ac{ev} has limited flexibility because it has to be charged before departure, the \ac{bess} is used to arbitrage energy following the frequencies of the $\lambda^{\textrm{IDA/CT}}_t$. On the heat side, the \ac{tess} delivers power until it gets close to $\underline{T}_{\textrm{TESS}}$ and struggles to be an attractive short-term flexibility provider when compared to electrical storage and under the series setup assumed in this work. Moreover, not all policies $\pi$ ensure a decrease in $C_{\text{grid}}$ when adding the \ac{tess}.

Lastly, when analyzing the effectiveness of the capacity fade control, it is clear that stacking markets in general increases the total capacity lost $Q_{\textrm{loss}}$. However, in most cases, moving from \ac{da} to \ac{ct} or \ac{ida} increases the quality of the degradation control by reducing the capacity lost per full equivalent cycle. Within these two groups of policies can be distinguished: $\pi_{DA2}$ and $\pi_{DA2IDA}$ focus on extending battery lifetime, whereas $\pi_{DA \rightarrow CT}$ and $\pi_{DA \rightarrow IDA}$ spend their degradation more effectively to improve the flexibility of the residential energy hub.

\subsection{Limitations \& Discussion}

Accurate forecasting of all \textbf{exogenous processes} ($W_{t+1}$) is a prerequisite for effective operation, which needs $B_{a,t} \approx W_{t+1}$. In particular, predicting the clearing price for the pay-as-clear \ac{ida} market is simpler than estimating the dynamic distribution of the pay-as-bid \ac{ct} market. This difficulty is compounded by the modelling assumption that the CT market functions as a \textbf{pay-as-clear auction} based on the ID1 index. Conversely, certain binary exogenous variables, such as \ac{ev} availability ($\gamma_{\textrm{EV},t}$) and building occupancy ($\mathcal{O}_t$), are often estimated directly from user input or pre-set schedules, offering a more readily available, albeit still imperfect, set of inputs for the \ac{ems}. Finally, weather forecasts can be fetched through APIs from KNMI, data retailers, or developed in-house.

For experimental implementation, the system needs state observers to accurately feed back the current device states ($S_{a,t}$) to the control policies ($\pi$). This is essential for the closed-loop functionality of the \ac{ems}. Furthermore, several simplifying assumptions were made regarding the thermal behavior of energy storage devices. Specifically, both the \ac{bess} and the \ac{ev} battery are assumed to maintain a constant operating temperature, implying HVAC systems for both. This assumption simplifies the control problem by neglecting the nonlinear effects of temperature on battery degradation and performance. In combination with the limited C-rates, Li-plating is also neglected, but future work should aim to incorporate temperature controls and additional battery degradation mechanisms, especially for countries with harsher weather conditions like snow and extreme heat.

Despite efforts to improve flexibility compared to standard literature, the terminal set design remains somewhat arbitrary. Future work should focus on designing the terminal conditions (Equations \ref{eq:pcDA} and \ref{eq:pcCT}) using a more rigorous and systematic criterion to enhance the stability and optimality of the long-term control horizon. More so when thinking that due to the short \ac{mpc} horizon the \ac{tess} flexibility is marginal, even hindering the overall flexibility as price volatility and test set heterogeneity grow. This trade-off between long-term operation, short-term flexibility and \ac{tess} appears as an interesting research direction.

As mentioned earlier, this work aims to challenge existing literature assumptions regarding the impact of sequential markets on energy hubs. This is manifested in the higher harmonic content of $\lambda^{\text{IDA/CT}}_t$ w.r.t $\lambda^{\textrm{DA}}_t$, which results in increased $Q_{\textrm{loss}}$, something previously not analyzed in the literature. Batteries exposed to sequential market operations will require adequate mitigation measures to control their degradation, driving the need for innovative business models not only for individual homes but for aggregated schemes. This type of consideration wouldn't be possible without the use of advanced control-oriented battery \ac{pbrom} \cite{VegaGarita2025TheReview}. Their adoption in more applications is a critical step for the massive adoption of battery technologies to assess revenues, warranties, and lifetimes with a broader lens.

Of course, it is highly unlikely that a single house has all the devices necessary for \textit{fullFlex} and with current market structures the business case for a single household is quite challenging. However, this paper shows that without specialized cloud-infrastructure or prohibitive computation hardware a \textit{fullFlex} ageing-aware \ac{empc} can be implemented in remote-servers in under $\Delta t^{\textrm{MPC}}=$15min since the mean computational time is $\mu_{\Delta}<2$min. Thus, the potential for aggregators to optimize behind the meter assets is significant. Moreover a roadmap for gradual implementation is given, since the short-term flexibility contribution is presented with the logical steps being \textit{noflex} $\rightarrow$ \textit{eFlex} $\rightarrow$ \textit{fullFlex}. On the infrastructure side, communication can be implemented over traditional protocols such as Modbus TCP or specialized data layers like S2 or OpenADR \cite{s2protocol}, depending on the available engineering hours.

Overall, the variation between the different policies $\pi$ are the price signals that they follow. As the volatility of such signals increases ($DA \rightarrow IDA \rightarrow CT$) in the short term the electrical storage is best suited to exploit it. To integrate the \ac{tess} into the \textit{eFlex} setup and become \textit{fullFlex} the volatility and heterogeneity of $W_{t+1}$ has to be reduced or break-down by decomposition techniques, special-warm starts or new thermal setup designs.

\section{Conclusions}
\label{sec:conclusions}


In summary, this paper presents a two-stage economic model predictive controller for residential energy hubs. The \ac{empc} can actively control battery ageing and thermal comfort through detailed physics-based models, while optimizing grid cost and charging the \ac{ev}. The presented formulation can be integrated into day-ahead and intra-day markets (auctions and continuous-time). Not only optimizing the day-ahead market, but also sequentially optimizing multiple markets. 

Focusing on the grid cost $C_{\text{grid}}$, our analysis shows that under sequential markets, the worst policy is to follow only the day-ahead prices $\pi_{DA2}$ or try to anticipate the \ac{ida} with $\pi_{DA2IDA}$ if the operating conditions $W_{t+1}$ are too heterogeneous. On the contrary, the best policy is optimizing for day-ahead and intra-day auction prices the \ac{bess} and \ac{ev}. During winter, or moments when $W_{t+1}$ resembles winter, it is best following day-ahead and then any intra-day market ($\pi_{DA\rightarrow IDA/CT}$).  For summer, it is best to follow the continuous time intra-day instead of the auction ($\pi_{DA\rightarrow CT}$) which exploits the duck curves of $\lambda^{\text{DA/CT}}_t$.  Unfortunately, integrating \ac{tess} with \ac{bess} and \ac{ev} under \ac{ct} hinders the flexibility of the energy hub. These contradict the common literature assumption that always following $\lambda^{\text{CT}}_t$ is the best possible policy.  If the focus is on extending battery lifetime, following the intra-day auction ($\pi_{DA \rightarrow IDA}$) is the top-performing policy, because it achieves the least possible capacity fade. Moreover, preliminary degradation analysis where only \ac{da} market participation is considered does not reflect the effective degradation achieved during implementation under sequential energy markets, since the full equivalent cycles of the batteries increase from \ac{da} to \ac{ct} or \ac{ida}.

Regarding the flexibility of the setup, the first case study presents the limited synergies between the heat and power storage. Our findings show that the integration of the \ac{tess} with the \ac{bess} and \ac{ev} unlocks additional savings mainly in the day-ahead market. However, \ac{tess} flexibility is only delivered under specific policies and input conditions. Only in summer with the \ac{ida} policy $\pi_{DA \rightarrow IDA}$. However, its total short-term realized flexibility is marginal when compared to \ac{bess} and \ac{ev} with bi-directional charging. The limiting factors are the assumed thermal piping, the \ac{tess} low round-trip efficiency and low C-rate. This is inline with current literature \cite{Alpizar-Castillo2024ModellingHouse, Darivianakis2017AManagement}. The incorporation of the \ac{tess} impacts battery degradation differently depending on the season. In the winter, the capacity lost increases for both \ac{bess} and \ac{ev}. For both, the addition of the \ac{tess} hinders the ageing control (increases $\frac{\partial Q_{\text{loss}}}{\partial FEC}$). 

In general, participating in sequential markets increases the overall degradation of the batteries de to the higher volatility of the intra-day markets, which is usually overlooked in the literature and by the manufacturers. Degradation controls are crucial to mitigate this additional capacity fade. In this line, the sequential policies $\pi_{DA \rightarrow CT}$ and $\pi_{DA \rightarrow IDA}$ are the most effective at converting lost battery capacity into short-term flexibility.

Future works aim at integrating the presented policies with local real-time controls and observers. A seasonal-planning layer to improve the coordination with the heat carrier and avoid the early depletion of the thermal storage is also attractive. Another direction is the explicit integration of exogenous uncertainty $W_{t+1}$ into the policy design. Finally, the addition of other relevant markets, such as frequency reserves and similar, also appear as attractive future research directions.

\section{Acknowledgment}

The project was carried out with a Top Sector Energy subsidy from the Ministry of Economic Affairs and Climate, carried out by the Netherlands Enterprise Agency (RVO). The specific subsidy for this project concerns the MOOI subsidy round 2020.

\appendix
\section{Parameters and models}
\label{sec:appA}

\begin{table}[th!]
    \centering
    \caption{Parameter values for the Building}
    \label{tab:building_data}
    \begin{tabular}{l cccc}
        \hline
        Parameter & Symbol  &Unit 
& \multicolumn{2}{c}{Value}\\
 &  &
& Winter& Summer\\
        \hline
        Air capacity & $C_{\text{air}}$&$\frac{\text{kWh}}{\text{kg.K}}$& \multicolumn{2}{c}{$0.279 \times 10^{-3}$}\\
        Air density & $\rho_{\text{air}}$&$\frac{\text{kg}}{\text{m}^3}$& \multicolumn{2}{c}{$1.225$}\\
        Building thermal capacity & $C_b$  &$\frac{\text{kWh}}{\text{K}}$& \multicolumn{2}{c}{$4.755$}\\
        Building volume & $V_b$  &m$^3$ 
& \multicolumn{2}{c}{$585$}\\
        Windows solar heat gain coefficient& $s_b$  &- 
& $0.5$  &0.1\\
        Building wall-to-wall ratio & $w_b$  &- 
& \multicolumn{2}{c}{$0.3$}\\
        Air change rate & $r_b$  &$\frac{1}{\text{h}}$& $0.35$  &0.99\\
        Thickness of the surfaces & $d$  &m 
& \multicolumn{2}{c}{$[0.03, 0.23, 0.23, 0.015]$}\\
        Conductivity of the surfaces & $U$  &$\frac{\text{kW}}{\text{m.K}}$& \multicolumn{2}{c}{$[0.18, 1., 1., 0.72] \times 10^{-3}$}\\
        Area of the surfaces & $A$  &m$^2$ 
& \multicolumn{2}{c}{$[90., 75., 48., 63.75]$}\\
        Mass flow of the fluid & $\dot{m}_f$&$\frac{\text{kg}}{\text{s}}$& \multicolumn{2}{c}{$0.22$}\\
        Specific heat capacity of the fluid & $c_f$  &$\frac{\text{kWh}}{\text{kg.K}}$& \multicolumn{2}{c}{$1.16 \times 10^{-3}$}\\
 HP Heat-exchanger thermal efficiency& $\eta_{\text{HP}}$& -& \multicolumn{2}{c}{0.8}\\
        Supply temperature setpoint & $T_{\text{sup}}$&K & \multicolumn{2}{c}{$323$}\\
 Building temperature bounds& $\underline{T}_{\text{in}},\ \overline{T}_{\text{in}}$& K & \multicolumn{2}{c}{[290, 297]}\\
 \hline
    \end{tabular}
\end{table}

\subsection{Thermal models}
Building model
\begin{subequations}
\begin{align}
    & T_{\text{in},t+1}  = T_{\textrm{in},t}+\frac{\Delta t}{C_b + V_b . \rho_{\textrm{air}} . C_{\textrm{air}}}  \left( \dot{Q}_{\textrm{ir},t} + \dot{Q}_{\textrm{TESS},t}^{\textrm{D}} + \dot{Q}_{\textrm{HP},t}^{\textrm{D}} - \dot{Q}_{\textrm{loss},t} \right)\\
    & \dot{Q}_{\textrm{ir},t}= w_b . s_b . G_{\textrm{ir},t} . \sum_{s=2}^3 A_s\\
    & \dot{Q}_{\textrm{loss},t}= \dot{Q}_{\textrm{cond},t}+\dot{Q}_{\textrm{vent},t} \\
    & \dot{Q}_{\textrm{vent},t}= C_{\textrm{air},t} . \rho_{\textrm{air}} .V_b .r_b . (T_{\text{in},t} - T_{\text{amb},t}) \\
    & \dot{Q}_{\textrm{cond},t}= (T_{\text{in},t} - T_{\text{amb},t}) \sum_{s=1}^{\mathcal{S}} U_s.A_s
\end{align}
\end{subequations}
TESS self-discharge $\dot{Q}_{\text{sd}}=1\%$ of the capacity $Q_{\text{TESS}}$.

\subsection{Electrical models}
Battery models
\begin{subequations}
    \begin{align}
        & SoC_{b,t+1}=SoC_{b,t}-\frac{\Delta t}{Q_{b,t}.3600}.\eta_c.i_{b,t} \\
        & P_{b,t}= N_{s,b}.N_{p,b}. v_{t,b,t} . i_{b,t} \\
        & i_{R_1,b,t+1}=e^{-\frac{\Delta t}{R_{1,b}.C_{1,b}}}.i_{R_1,b,t}+ \left( 1-e^{-\frac{\Delta t}{R_{1,b}.C_{1,b}}} \right) .i_{b,t} \\
        & OCV_{b,t} = OCV_{p,b,t}(SoC_{b,t}) - OCV_{n,b,t}(SoC_{b,t})\\
        & v_{t,b,t}=OCV_{b,t}-i_{R_1,b,t}.R_{1,b}-i_{b,t}.R_{0,b} \\
        & z_{b,t} = SoC_{b,t}.(z_{100\%}-z_{0\%})+z_{0\%} \\
        & \eta_{k,b,t}=\frac{2.R.T}{F}.\text{sinh}^{-1}\left(\frac{i_{b,t}}{n_{\textrm{SEI}}.a_s.A.L_n.i_0}\right) \\
        & \beta_{b,t} = e^{\frac{n_{\textrm{SEI}}.F}{R.T}.\left(\eta_{k,b,t}+OCV_{n,b,t}-OCV_s \right)} \\
        & i_{\textrm{SEI},b,t} = \frac{k_{\textrm{SEI},b}.e^{\frac{-E_{\textrm{SEI},b}}{RT}}}{n_{\textrm{SEI}}.(1+\lambda_{b}.\beta_{b,t}).\sqrt{t}} \\
        & i_{\textrm{AM},b,t} = k_{\textrm{AM},b}.e^{\frac{-E_{\textrm{AM},b}}{R.T}}.SoC_{b,t}.|i_{b,t}|.Q_{b,0}\\
        & i_{\text{loss},\ b,t} = i_{\textrm{SEI},b,t} + i_{\textrm{AM},b,t} \\
        & Q_{b,t+1}=Q_{b,t} -\frac{\Delta t}{3600}.i_{\text{loss},b,t} \\
        & S_{b,t}=\left[SoC,\ i_{R_1},\ OCV_{j},\ z,\ \eta_{k},\ \beta,\ i_{\text{SEI}},\ i_{\text{AM}} \right]^T_{b,t}\\
        & x_{b,t} = P_{b,t},\ y_{b,t} = \left[i, v_t \right]^T_{b,t},\ \theta_{b,t}=Q_{b,t}
    \end{align}
\end{subequations}

Electric vehicle
\begin{subequations}
    \begin{align}
        & \gamma_t=\begin{cases}
          0 & t \in [t_{\text{dep}};\ t_{\text{arr}}]\\
          1 & \text{otherwise}
        \end{cases} \,. \\
        & P_{\textrm{tot},\textrm{EV},t}=\gamma_{\textrm{EV},t}.P_{\textrm{EV},t}+(1-\gamma_{\textrm{EV},t})P_{\textrm{drive},\textrm{EV}}\\
        & \varepsilon_{SoC}=SoC_{\textrm{EV}}(t_{\textrm{dep}})-SoC_{\textrm{dep}}^*\\
        & p_{\textrm{SoCDep}}=w_{SoC}.||\varepsilon_{SoC}||_2^2
    \end{align}
\end{subequations}

\section{Extended results}
\label{sec:appB}
\begin{figure}[h!]
    \centering
    \includegraphics[width=0.9\columnwidth]{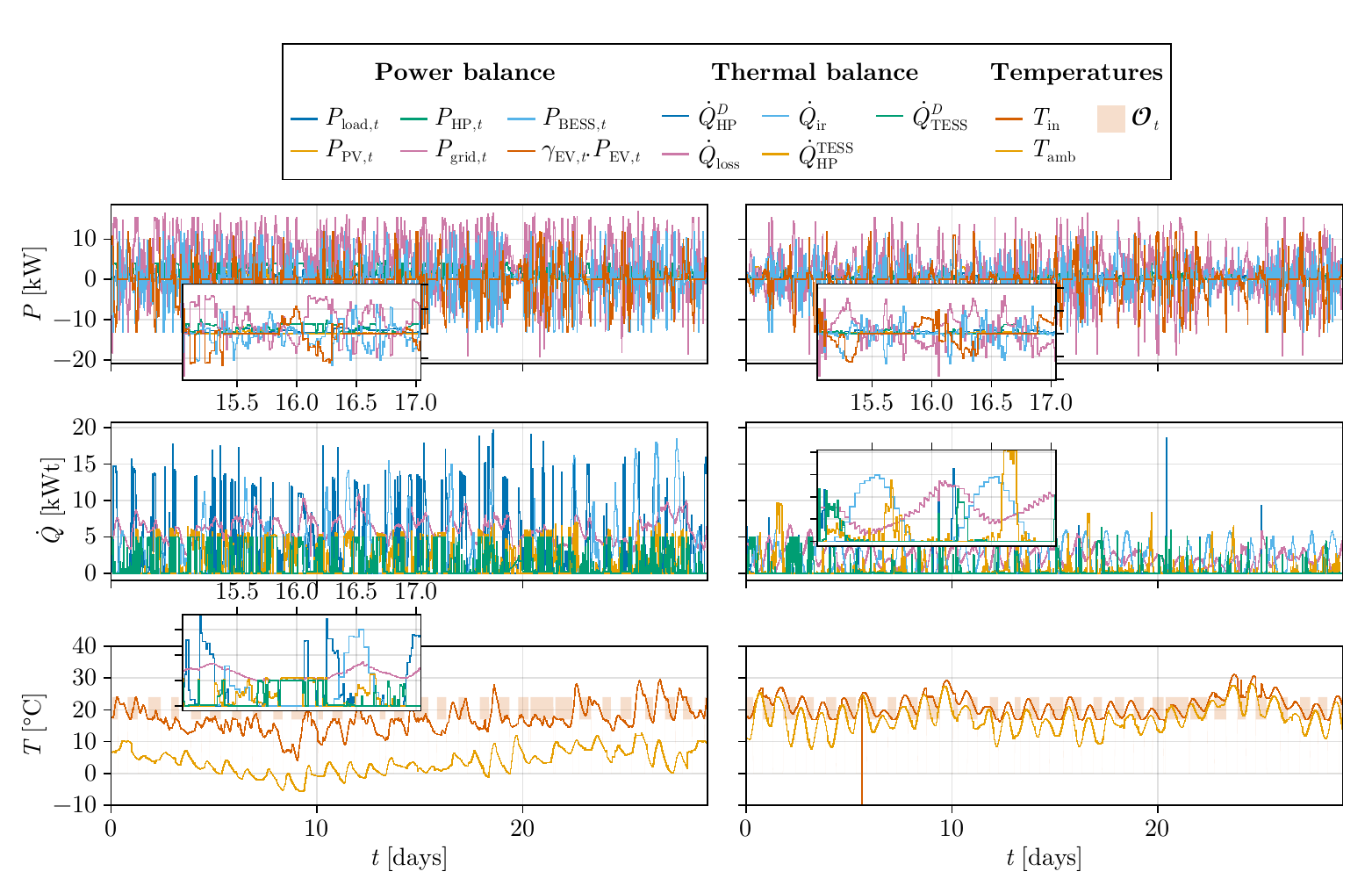}
    \caption{Summary of the $\pi_{DA \rightarrow IDA}$ monthly simulations, (left ) winter and (right) summer. a) Power Balance. b) Heat balance. d) Building temperatures.}
    \label{fig:cs0_ems_fullFlex_daida}
\end{figure}

\begin{figure}[h!]
    \centering
    \includegraphics[width=0.9\columnwidth]{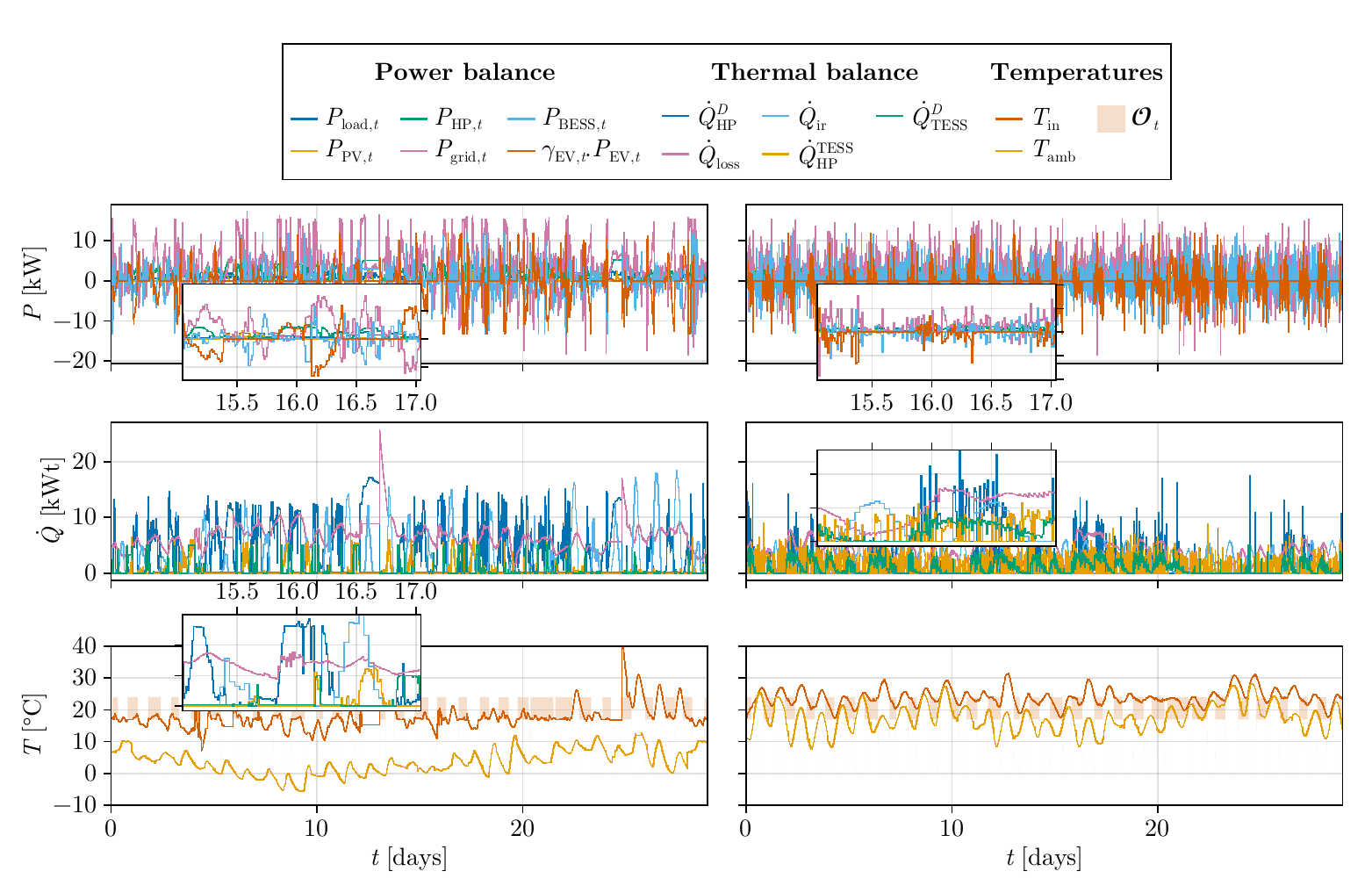}
    \caption{Summary of the $\pi_{DA \rightarrow CT}$ for weekly simulations, with the left column being summer and right column being winter. a) Power Balance. b) Heat balance. d) Building temperatures.}
    \label{fig:cs0_ems_fullFlex_dact}
\end{figure}

\begin{figure}[h!]
    \centering
    \includegraphics[width=0.9\columnwidth]{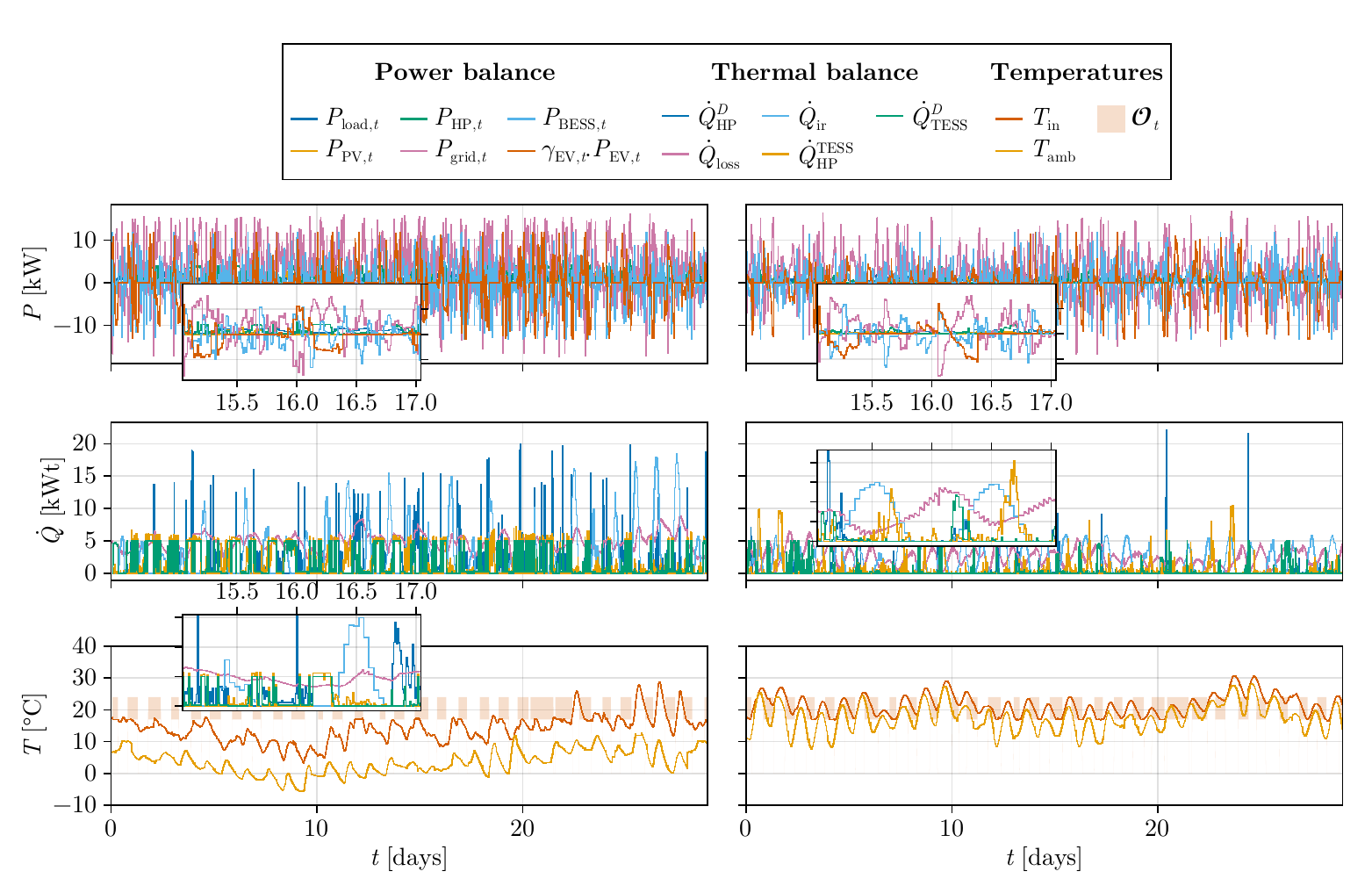}
    \caption{Summary of the $\pi_{DA +IDA\rightarrow IDA}$ for weekly simulations, with the left column being summer and right column being winter. a) Power Balance. b) Heat balance. d) Building temperatures.}
    \label{fig:cs0_ems_fullFlex_da2ida}
\end{figure}

The Figures \ref{fig:cs0_ems_fullFlex_da2} - \ref{fig:cs0_ems_fullFlex_da2ida} present the rest of the \textit{fullFlex} balances and inside temperature for the rest of the policies $\pi_{DA\rightarrow IDA}$, $\pi_{DA \rightarrow CT}$ and $\pi_{DA2IDA}$.

\begin{figure}[h!]
    \centering
    \includegraphics[width=\textwidth]{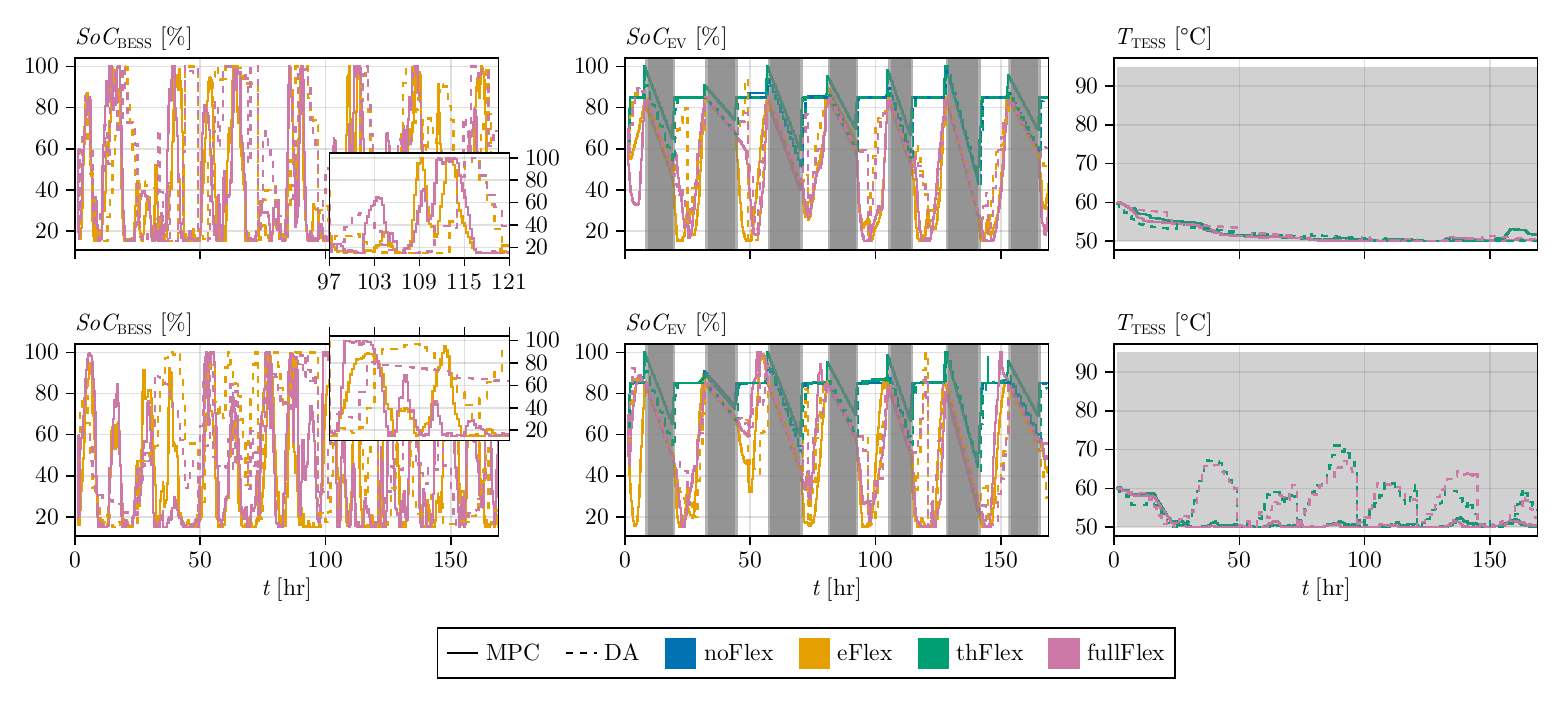}
    \caption{HESS states under the $\pi_{2DA}$ in summer (top) winter (bottom).}
    \label{fig:cs1_hess_da2}
\end{figure}

\begin{figure}[h!]
    \centering
    \includegraphics[width=\textwidth]{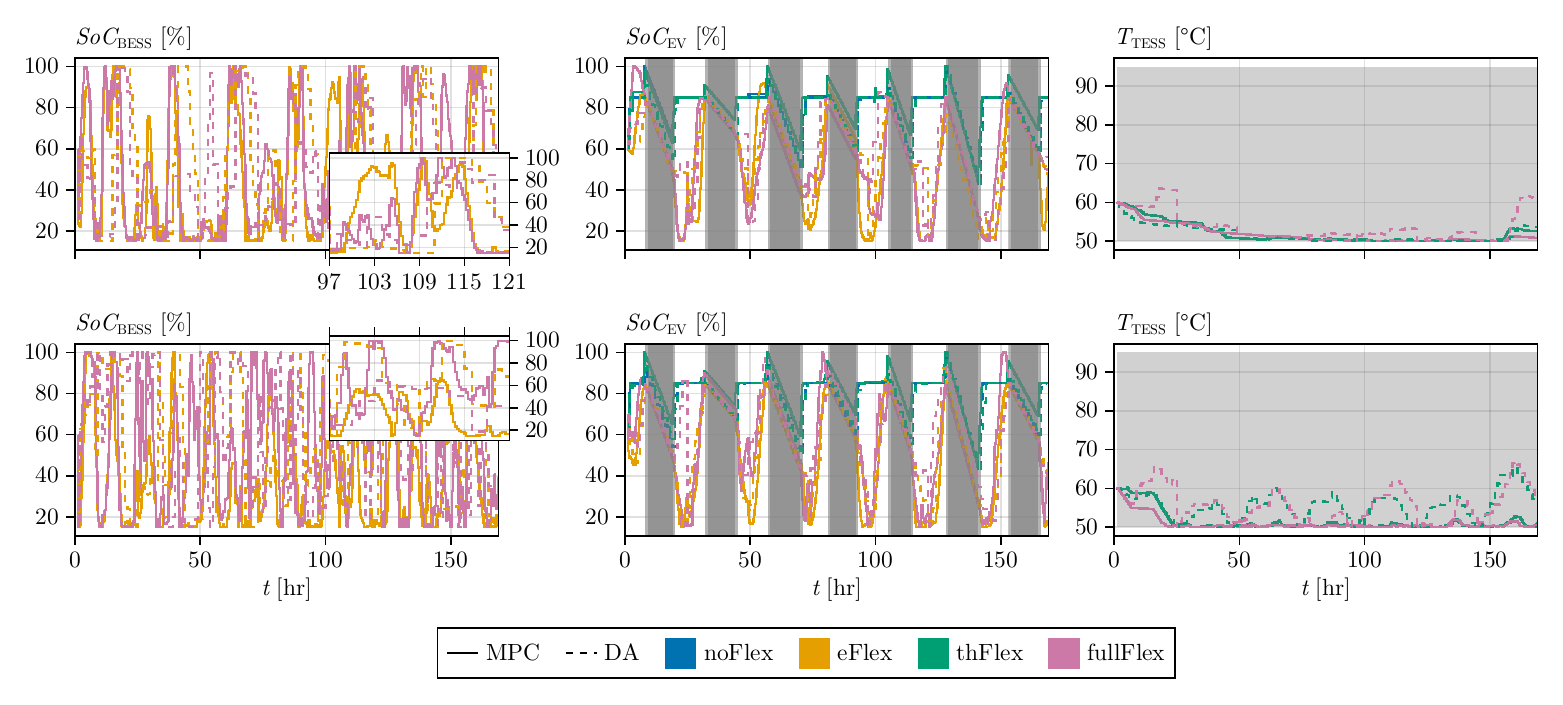}
    \caption{HESS states under the $\pi_{DA\rightarrow CT}$ in summer (top) winter (bottom).}
    \label{fig:cs1_hess_dact}
\end{figure}

\begin{figure}[h!]
    \centering
    \includegraphics[width=\textwidth]{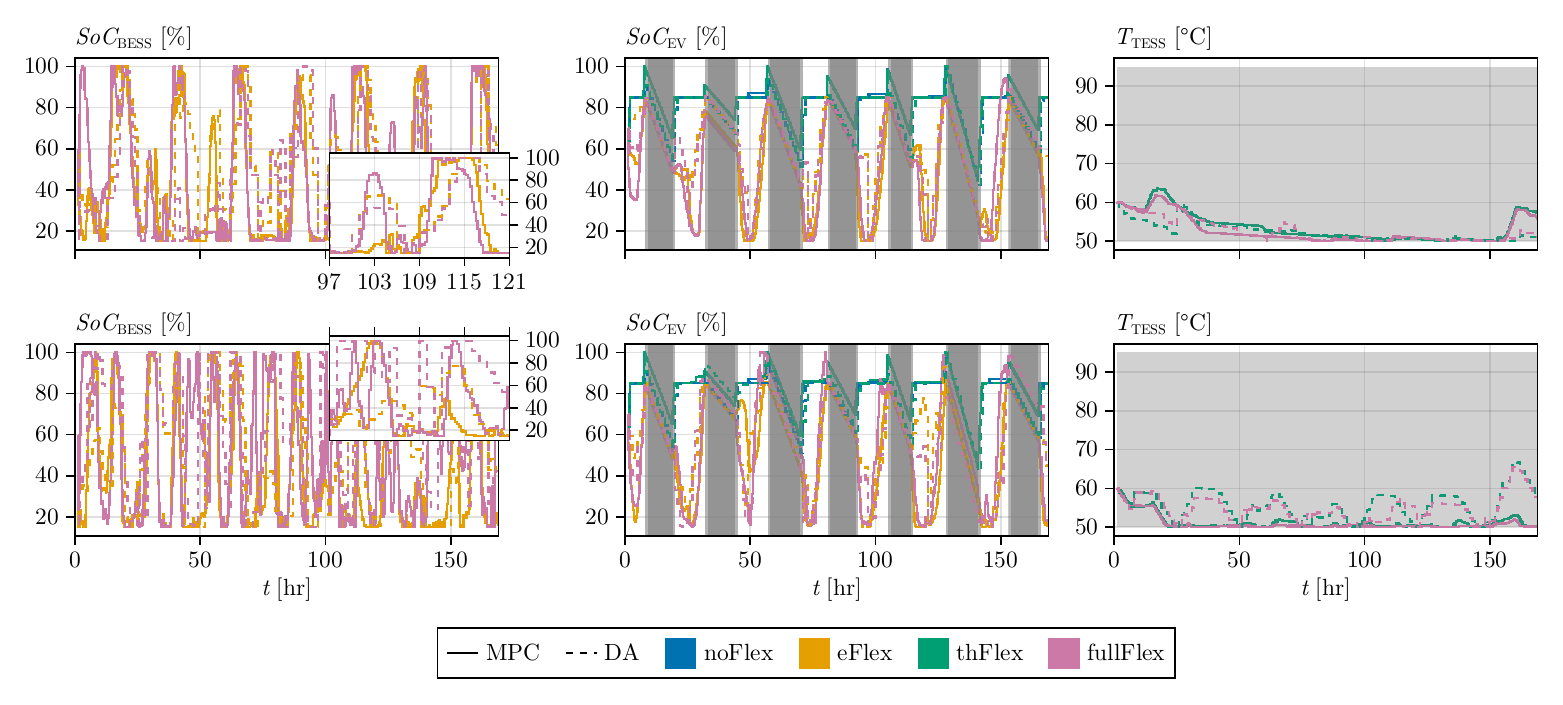}
    \caption{HESS states under the $\pi_{DA2IDA}$ in summer (top) winter (bottom).}
    \label{fig:cs1_hess_da2ida}
\end{figure}

The Figures \ref{fig:cs1_hess_da2} - \ref{fig:cs1_hess_da2ida} present the rest of the \ac{hess} operation plots the rest of the policies $\pi_{DA2}$, $\pi_{DA \rightarrow CT}$ and $\pi_{DA2IDA}$ for standard weeks $T=1$ week.
\begin{figure}
    \centering
    \includegraphics[width=0.5\linewidth]{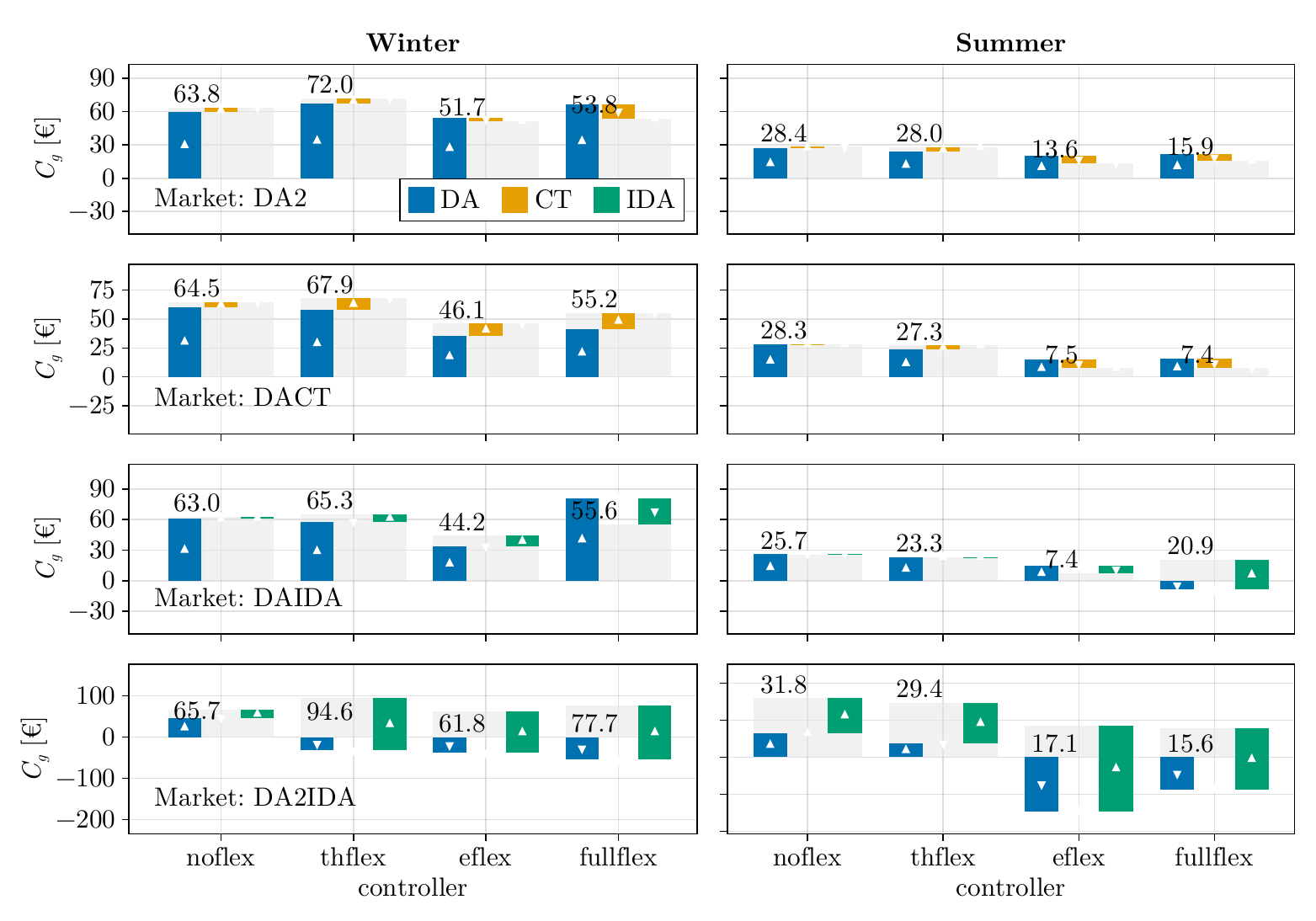}
    \caption{Grid flexibility provided per setup and market. For winter (left), and summer (right) and weekly standard weeks.}
    \label{fig:cs1_compareCgrid_d7}
\end{figure}


Table \ref{tab:compTimes} presents a detailed breakdown per policy and per flexibility setup.

\begin{table}[bt]
\centering
\begin{adjustbox}{width=1.15\textwidth}
    \begin{tabular}{l l ll ll ll ll ll ll lll}\hline
     & \multicolumn{8}{c}{Winter}&\multicolumn{8}{c}{Summer}\\
    case & \multicolumn{2}{c}{DA2}&\multicolumn{2}{c}{DACT}&\multicolumn{2}{c}{DAIDA}&\multicolumn{2}{c}{DA2IDA}&\multicolumn{2}{c}{DA2}&\multicolumn{2}{c}{DACT}&\multicolumn{2}{c}{DAIDA}&\multicolumn{2}{c}{DA2IDA}\\
     & $\mu$& $\sigma$& $\mu$& $\sigma$& $\mu$& $\sigma$& $\mu$& $\sigma$& $\mu$& $\sigma$& $\mu$& $\sigma$& $\mu$& $\sigma$& $\mu$&$\sigma$\\
     \hline
    noflex & 5,85  &   13,84 
    &18,024&  33,1184&32,98  &   63,40 
    &6,26  &   14,77 
    &13,03  &   28,78 
    &-  &  &20,00  &   41,38 
    &14,20   & 31,83 
    \\
    thflex & 12,03  &   22,38 
    &-&  &7,42  &   15,30 
    &12,08  &   21,91 
    &10,20  &   20,89 
    &-  &  &8,85  &   17,87 
    &8,65   & 17,57 
    \\
    eFlex & 5,97  &   15,87 
    &-&  &6,11  &   11,96 
    &12,70  &   15,24 
    &9,61  &   18,88 
    &-&  &10,99  &   24,48 
    &11,61   & 15,36 
    \\
    fullFlex & 25,92  &   89,85 
    &-  &  &16,07  &   85,53 
    &3,76  &   10,04 
    &25,55  &   56,56 
    &17,28&  45,9192&47,68  &   131,70 
    &2,75   & 3,92 
    \\ \hline
    \end{tabular}
\end{adjustbox}
\caption{Mean and standard deviation of the computational times $\Delta_{\text{comp}}$.}
\label{tab:compTimes}
\end{table}

\printglossaries
\printnomenclature

 \bibliographystyle{elsarticle-num} 
 \bibliography{references}
 \biboptions{sort&compress}





\end{document}